%% file: ms.tex
\documentclass[12pt]{article}
\usepackage{amsmath,amssymb}
\usepackage{bbm}
\usepackage{booktabs}
\usepackage{ntheorem}
\usepackage{bigints}
\usepackage{colonequals}
\usepackage[official]{eurosym}
\usepackage[margin=1in]{geometry}
\usepackage{enumerate}
\usepackage[round]{natbib}
\usepackage[onehalfspacing]{setspace}
\usepackage{color,graphicx,subcaption,hyperref}
\usepackage{rotating}
\usepackage{dutchcal}
\hypersetup{pdfborder = {0 0 0},colorlinks=true,linkcolor=blue,urlcolor=blue,citecolor=blue}

\usepackage[mathscr]{euscript} 

\usepackage{xr}

\newtheorem{assumption}{Assumption}

\newtheorem{definition}{Definition}

\newtheorem{theorem}{Theorem}

\renewcommand{\L}{\mathcal{l}}
\renewcommand{\H}{\mathcal{h}}
\newcommand{\x}{\bar{x}}
\newcommand{\1}{\mathbbm{1}}

\DeclareMathOperator*{\argmin}{argmin}

\newcommand{\E}{\mathbbm{E}}
\newcommand{\V}{\mathbbm{V}}
\newcommand{\cov}{\mathbbm{C}\mathrm{ov}}

\newcommand{\pathin}{./inputs}

\begin{document}
	
\title{Extrapolating Treatment Effects in Multi-Cutoff Regression Discontinuity Designs\thanks{We are very grateful to Fabio Sanchez and Tatiana Velasco for sharing the dataset used in the empirical application. We also thank Josh Angrist, Sebastian Calonico, Sebastian Galiani, Nicolas Idrobo, Xinwei Ma, Max Farrell, and seminar participants at various institutions for their comments. We also thank the co-editor, Regina Liu, an associate editor and a reviewer for their comments. Cattaneo and Titiunik gratefully acknowledges financial support from the National Science Foundation (SES 1357561).}}

\author{
	Matias D. Cattaneo\footnote{Department of Operations Research and Financial Engineering, Princeton University.}\and
	Luke Keele\footnote{Department of Surgery and Biostatistics, University of Pennsylvania.}\and
	Roc\'{i}o Titiunik\footnote{Department of Politics, Princeton University.} \and
	Gonzalo Vazquez-Bare\footnote{Department of Economics, University of California at Santa Barbara.}
}
\maketitle
\setcounter{page}{0}\thispagestyle{empty}

\begin{abstract}
	In non-experimental settings, the Regression Discontinuity (RD) design is one of the most credible identification strategies for program evaluation and causal inference. However, RD treatment effect estimands are necessarily local, making statistical methods for the extrapolation of these effects a key area for development. We introduce a new method for extrapolation of RD effects that relies on the presence of multiple cutoffs, and is therefore design-based. Our approach employs an easy-to-interpret identifying assumption that mimics the idea of ``common trends'' in difference-in-differences designs. We illustrate our methods with data on a subsidized loan program on post-education attendance in Colombia, and offer new evidence on program effects for students with test scores away from the cutoff that determined program eligibility.
\end{abstract}

	
\doublespacing\newpage
\setlength{\abovedisplayskip}{5pt}
\setlength{\belowdisplayskip}{5pt}

\section{Introduction}\label{sec:intro}

The regression discontinuity (RD) design is one of the most credible strategies for estimating causal treatment effects in non-experimental settings. In an RD design, units receive a score (or running variable), and a treatment is assigned based on whether the score exceeds a known cutoff value: units with scores above the cutoff are assigned to the treatment condition, and units with scores below the cutoff are assigned to the control condition. This treatment assignment rule creates a discontinuity in the probability of receiving treatment which, under the assumption that units' average characteristics do not change abruptly at the cutoff, offers a way to learn about the causal treatment effect by comparing units barely above and barely below the cutoff. Despite the popularity and widespread use of RD designs, the evidence they provide has an important limitation: the RD causal effect is only identified for the very specific subset of the population whose scores are ``just'' above or below the cutoff, and is not necessarily informative or representative of what the treatment effect would be for units whose scores are far from the RD cutoff. Thus, by its very nature, the RD parameter is local and has limited external validity.

We empirically illustrate the advantages and limitations of RD designs employing a recent study of the ACCES (\textit{Acceso con Calidad a la Educaci\'on Superior}) program by \citet{MelguizoSanchezVelasco2016-WD}. ACCES is a subsidized loan program in Colombia, administered by the Colombian Institute for Educational Loans and Studies Abroad (ICETEX), that provides tuition credits to underprivileged populations for various post-secondary education programs such as technical, technical-professional, and university degrees. In order to be eligible for an ACCES credit, students must be admitted to a qualifying higher education program, have good credit standing and, if soliciting the credit in the first or second semester of the higher education program, achieve a minimum score on a high school exit exam known as \textit{SABER 11}. In other words, to obtain ACCES funding students must have an exam score above a known cutoff. Students who are just below the exam cutoff are deemed ineligible, and therefore are not offered financial assistance. This discontinuity in program eligibility based on the exam score leads to a RD design: \citet{MelguizoSanchezVelasco2016-WD} found that students just above the threshold in SABER 11 test scores were significantly more likely to enroll in a wide variety of post-secondary education programs. The evidence from the original study is limited to the population of students around the cutoff. This standard causal RD treatment effect is informative in its own right but, in the absence of additional assumptions, it cannot be used to understand the effects of the policy for students whose test scores are outside the immediate neighborhood of the cutoff. Treatment effects away from the cutoff are useful for a variety of purposes, ranging from answering purely substantive questions to addressing practically important policy making decisions such as whether to roll-out the program or not.

We propose a novel approach for estimating RD causal treatment effects away from the cutoff that determines treatment assignment. Our extrapolation approach is design-based as it exploits the presence of multiple RD cutoffs across different subpopulations to construct valid counterfactual extrapolations of the expected outcome of interest, given different scores levels, in the absence of treatment assignment. In sum, our approach imputes the average outcome in the absence of treatment of a treated subpopulation exposed to a given cutoff, using the average outcome of another subpopulation exposed to a higher cutoff. Assuming that the difference between these two average outcomes is constant as a function of the score, this imputation identifies causal treatment effects at score values higher than the lower cutoff. 

The rest of the article is organized as follows. The next section presents further details on the operation of the ACCES program, discusses the particular program design features that we use for the extrapolation of RD effects, and presents the intuitive idea behind our approach. In that section, we also discuss related literature on RD extrapolation as well as on estimation and inference. Section \ref{sec:methods} presents the main methodological framework and extrapolation results for the case of the ``Sharp'' RD design, which assumes perfect compliance with treatment assignment (or a focus on an intention-to-treat parameter). Section \ref{sec:analysis} applies our results to extrapolate the effect of the ACCES program on educational outcomes, while Section \ref{sec:simuls} illustrates our methods using simulated data. Section \ref{sec:extensions} presents an extension to the ``Fuzzy'' RD design, which allows for imperfect compliance. Section \ref{sec:conclusion} concludes. The supplemental appendix contains additional results, including further extensions and generalizations of our extrapolation methods.

\section{The RD Design in the ACCES Program}
 
The SABER 11 exam that serves as the basis for eligibility to the ACCES program is a national exam administered by the Colombian Institute for the Promotion of Postsecondary Education (ICFES), an institute within Colombia's National Ministry of Education. This exam may be taken in the fall or spring semester each year, and has a common core of mandatory questions in seven subjects---chemistry, physics, biology, social sciences, philosophy, mathematics, and language. To sort students according to their performance in the exam, ICFES creates an index based on the difference between (i) a weighted average of the standardized grades obtained by the student in each common core subject, and (ii) the within-student standard deviation across the standardized grades in the common core subjects. This index is commonly referred to as the SABER 11 score.

Each semester of every year, ICFES calculates the 1,000-quantiles of the SABER 11 score among all students who took the exam that semester, and assigns a score between 1 and 1,000 to each student according to their position in the distribution---we refer to these scores as the SABER 11 position scores. Thus, the students in that year and semester whose scores are in the top 0.1\% are assigned a value of 1 (first position), the students whose scores are between the top 0.1\%  and 0.2\% are assigned a value of 2 (second position), etc., and the students whose scores are in the bottom 0.1\% are assigned a value of 1,000 (the last position). Every year, the position scores are created separately for each semester, and then pooled. \citet{MelguizoSanchezVelasco2016-WD} provide further details on the Colombian education system and the ACCES program.

In this sharp RD design, the running variable is the SABER 11 position score, denoted by $X_i$ for each unit $i$ in the sample, and the treatment of interest is receiving approval of the ACCES credit. Between $2000$ and $2008$, the cutoff to qualify for an ACCES credit was $850$ in all Colombian departments (the largest subnational administrative unit in Colombia, equivalent to U.S. states). To be eligible for the program, a student must have a SABER 11 position score at or below the 850 cutoff.

\subsection{The Multi-Cutoff RD Design}\label{sec:multicutoffs}

In the canonical RD design, a single cutoff is used to decide which units are treated. As we noted above, eligibility for the ACCES program between $2000$ and $2008$ followed this template, since the cutoff was $850$ for all students. However, in many RD designs, the same treatment is given to all units based on whether the RD score exceeds a cutoff, but different units are exposed to different cutoffs. This contrasts with the assignment rule in the standard RD design, in which all units face the same cutoff value. RD designs with multiple cutoffs, which we call Multi-Cutoff RD designs, are fairly common and have specific properties \citep{Cattaneo-Keele-Titiunik-VazquezBare_2016_JOP}. 

In $2009$, ICFES changed the program eligibility rule, and started employing different cutoffs across years and departments. Consequently, after $2009$, ACCES eligibility follows a Multi-Cutoff RD design: the treatment is the same throughout Colombia---all students above the cutoff receive the same financial credits for educational spending---but the cutoff that determines  treatment assignment varies widely by department and changes each year, so that different sets of students face different cutoffs. This design feature is at the core of our approach for extrapolation of RD treatment effects.

\subsection{The Pooled RD Effect of the ACCES Program}\label{sec:pooled}

Multi-Cutoff RD designs are often analyzed as if they had a single cutoff. For example, in the original analysis, \citet{MelguizoSanchezVelasco2016-WD} redefined the RD running variable as distance to the cutoff, and analyzed all observations together using a common cutoff equal to zero. In fact, this normalizing-and-pooling approach \citep{Cattaneo-Keele-Titiunik-VazquezBare_2016_JOP}, which essentially ignores or ``averages over'' the multi-cutoff features of the design, is widespread in empirical work employing RD designs. See the supplemental appendix for a sample of recent papers that analyze RD designs with multiple cutoffs across various disciplines.

We first present some initial empirical results using the normalizing-and-pooling approach as a benchmark for later analyses. The outcome we analyze is an indicator for whether the student enrolls in a higher education program, one of several outcomes considered in the original study by \citet{MelguizoSanchezVelasco2016-WD}. In order to maintain the standard definition of RD assignment as having a score above the cutoff, we multiply the SABER 11 position score by $-1$. We focus on the intention-to-treat effect of program eligibility on higher education enrollment, which gives a Sharp RD design. We discuss an extension to Fuzzy RD designs in Section \ref{sec:extensions}. We focus our analysis on the population of students exposed to two different cutoffs, $-850$ and $-571$.

For our main analysis, we employ statistical methods for RD designs based on recent methodological developments in \citet{Calonico-Cattaneo-Titiunik_2014_ECMA,Calonico-Cattaneo-Titiunik_2015_JASA}, \citet{Calonico-Cattaneo-Farrell_2018_JASA,Calonico-Cattaneo-Farrell_2020_ECTJ,Calonico-Cattaneo-Farrell_2020_wp}, \citet{Calonico-Cattaneo-Farrell-Titiunik_2019_RESTAT}, and references therein. In particular, point estimators are constructed using mean squared error (MSE) optimal bandwidth selection, and confidence intervals are formed using robust bias correction (RBC). We provide details on estimation and inference in Section \ref{sec:estimation}. 

\begin{figure}
	\centering
	\caption{Normalizing-and-Pooling RD Plot of ACCES Loan Eligibility on Post-Education Enrollment.\label{fig:pld}}
	\begin{subfigure}[b]{0.45\textwidth}
		\includegraphics[width=\textwidth]{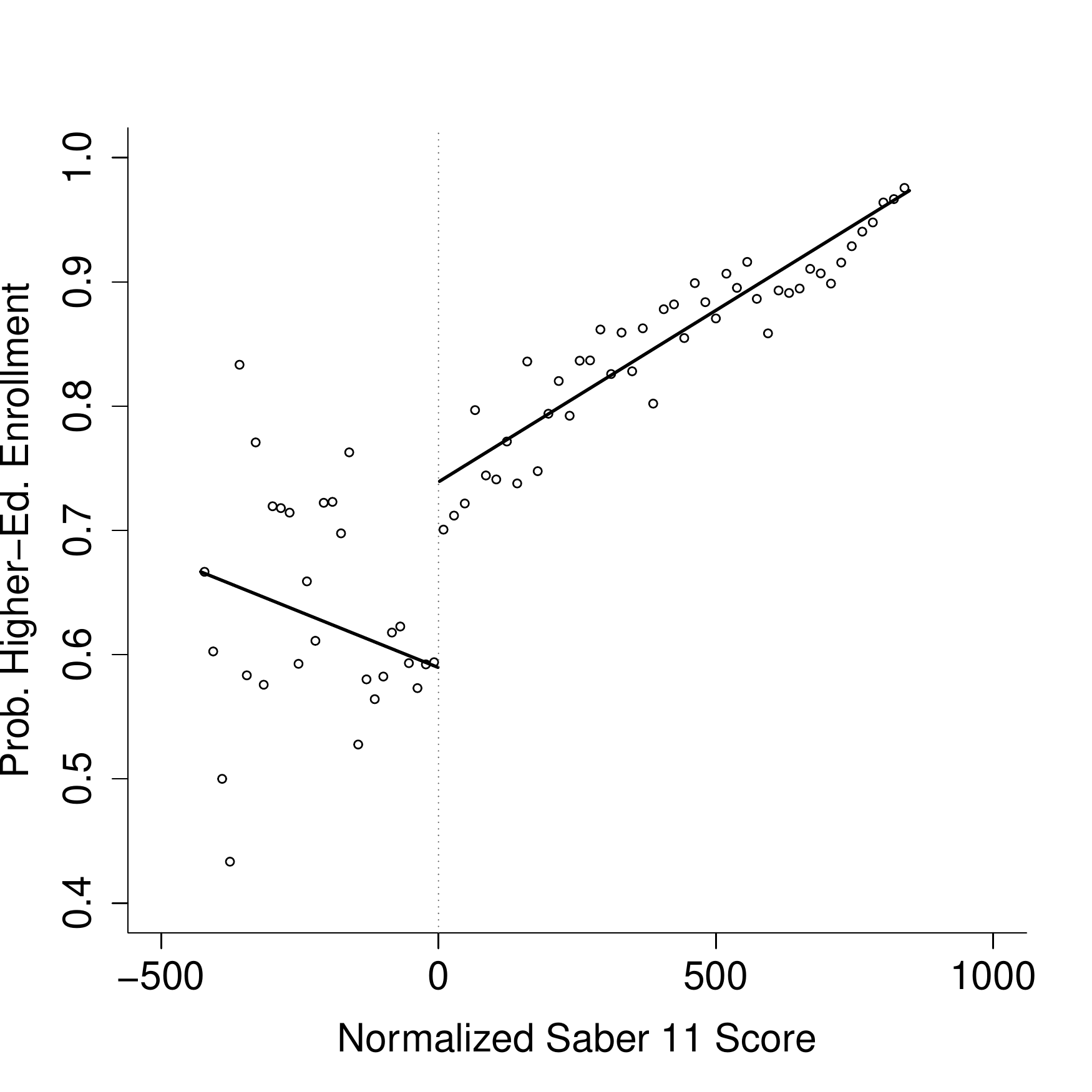}
		\caption{Global Linear Fit \label{fig:pld1}}
		\vspace{-.25in}
	\end{subfigure}
	\begin{subfigure}[b]{0.45\textwidth}
		\includegraphics[width=\textwidth]{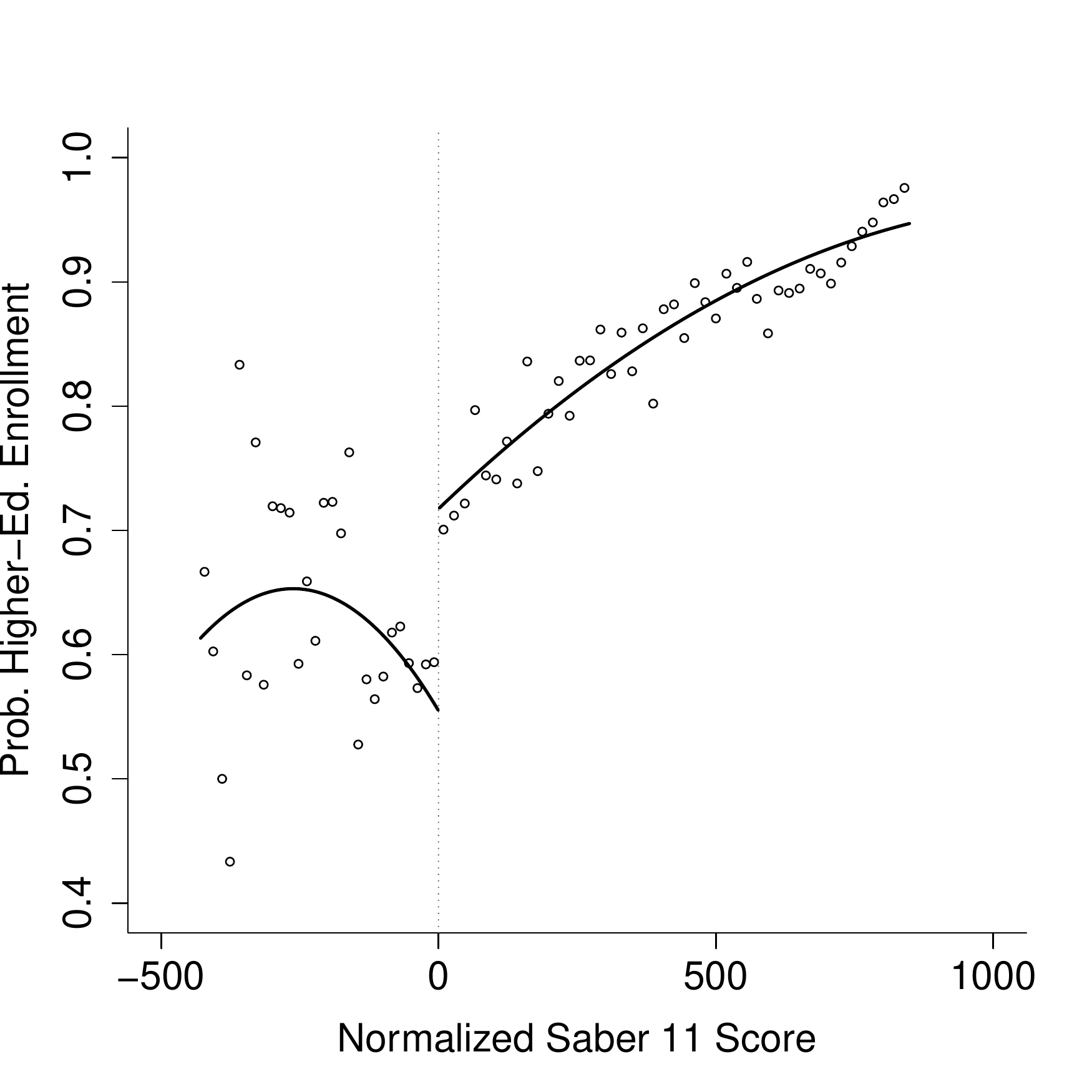}
		\caption{Global Quadratic Fit \label{fig:pld2}}
		\vspace{-.25in}
	\end{subfigure}
	
	\begin{flushleft}\footnotesize Notes. RD Plot constructed using evenly-spaced binning and global linear (left) and quadratic (right) polynomial fits for normalized (to zero) and pooled (across cutoffs) score variable. See \cite{Calonico-Cattaneo-Titiunik_2015_JASA} and \cite{Cattaneo-Keele-Titiunik-VazquezBare_2016_JOP} for methodological details, and \cite{Calonico-Cattaneo-Farrell-Titiunik_2017_Stata} and \cite{Cattaneo-Titiunik-VazquezBare_2020_Stata} for implementation. \end{flushleft}
\end{figure}

Figure \ref{fig:pld} reports two RD plots of the data, reporting linear and quadratic global polynomial approximations. Using RBC local polynomial inference, Table \ref{tab:results} reports that the pooled RD estimate of the ACCES program treatment effect on expected higher education enrollment is $0.125$, with corresponding $95\%$ RBC confidence interval $[0.012 , 0.219]$. These results indicate that, in our sample, students who barely qualify for the ACCES program based on their SABER 11 score are 12.5 percentage points more likely to enroll in a higher education program than students who are barely ineligible for the program. These results are consistent with the original positive effects of ACCES eligibility on higher education enrollment rates reported in \cite{MelguizoSanchezVelasco2016-WD}. However, the pooled RD estimate only pertains to a limited set of ACCES applicants: those whose scores are barely above or below one of the cutoffs.

\citet{Cattaneo-Keele-Titiunik-VazquezBare_2016_JOP} show that the pooled RD estimand is a weighted average of cutoff-specific RD treatment effects for subpopulations facing different cutoffs. The empirical results for the pooled and cutoff-specific estimates can be seen in the upper panel of Table \ref{tab:results}. In our sample, the pooled estimate of 0.125 is a (linear in large samples) combination of two cutoff-specific RD estimates, one for units facing the low cutoff $-850$ and one for units facing the high cutoff $-571$. We provide a detailed analysis of these estimates in Section \ref{sec:analysis}. These cutoff-specific estimates are not directly comparable, as these magnitudes correspond not only to different values of the running variable but also to different subpopulations. We discuss next how the availability of multiple cutoffs can be exploited to learn about treatment effects far from the cutoff in the context of the ACCES policy intervention.
 
\subsection{Using the Multi-Cutoff RD Design for Extrapolation}

Our key contribution is to exploit the presence of multiple RD cutoffs to extrapolate the standard RD average treatment effects (at each cutoff) to students whose SABER 11 scores are away from the cutoff actually used to determine program eligibility. Our method relies on a simple idea: when different units are exposed to different cutoffs, different units with the same value of the score may be assigned to different treatment conditions, relaxing the strict lack of overlap between treated and control scores that is characteristic of the single-cutoff RD design.

For example, consider the simplest Multi-Cutoff RD design with two cutoffs, $\L$ and $\H$, with $\L < \H$, where we wish to estimate the average treatment effect at a point $\x \in (\L,\H)$. Units exposed to $\L$ receive the treatment according to  $\mathbbm{1}(X_i\ge \L)$, where $X_i$ is unit's $i$ score  and $\mathbbm{1}(.)$ is the indicator function, so they are all treated at $X=\x$. However, the same design contains units who receive the treatment according to $\mathbbm{1}(X_i\ge \H)$, so they are controls at both $X=\x$ and $X=\L$. Our idea is to compare the observable difference in the control groups at the low cutoff $\L$, and assume that the same difference in control groups occurs at the interior point $\x$. This allows us to identify the average treatment effect for all score values between the cutoffs $\L$ and $\H$.

Our identifying idea is analogous to the ``parallel trends'' assumption in difference-in-difference designs \citep[see, e.g.,][and references therein]{Abadie_2005_RES}, but over a continuous dimension---that is, over the values of the continuous score variable $X_i$.

\subsection{Related Literature\label{sec:literature}}

We contribute to the causal inference and program evaluation literatures \citep{Imbens-Rubin_2015_Book,Abadie-Cattaneo_2018_ARE} and, more specifically, to the methodological literature on RD designs. See \cite{Imbens-Lemieux_2008_JoE}, \cite{Cattaneo-Titiunik-VazquezBare_2017_JPAM}, \cite{Cattaneo-Titiunik-VazquezBare_2020_Sage} and \cite{Cattaneo-Idrobo-Titiunik_2019_Book,Cattaneo-Idrobo-Titiunik_2020_Book}, for literature reviews, background references, and practical introductions.

Our paper adds to the recent literature on RD treatment effect extrapolation methods, a nonparametric identification problem in causal inference. This strand of the literature can be classified into two groups: strategies assuming the availability of external information, and strategies based only on information from within the research design. Approaches based on external information include \cite{Mealli-Rampichini_2012_JRSSA}, \cite{WingCook2013-JPAM}, \cite{Rokkanen2015-wp}, and \cite{AngristRokkanen2015-JASA}. The first two papers rely on a pre-intervention measure of the outcome variable, which they use to impute the treated-control differences of the post-intervention outcome above the cutoff. \cite{Rokkanen2015-wp} assumes that multiple measures of the running variable are available, and all measures capture the same latent factor; identification relies on the assumption that the potential outcomes are conditionally independent of the available measurements given the latent factor. \cite{AngristRokkanen2015-JASA} rely on pre-intervention covariates, assuming that the running variable is ignorable conditional on the covariates over the whole range of extrapolation. All these approaches assume the availability of external information that is not part of the original RD design.

In contrast, the extrapolation approaches in \cite{Dong-Lewbel_2015_ReStat} and \citet{Bertanha-Imbens_2020_JBES} require only the score and outcome in the standard (single-cutoff) RD design. \cite{Dong-Lewbel_2015_ReStat} assume mild smoothness conditions to identify the derivatives of the average treatment effect with respect to the score, which allows for a local extrapolation of the standard RD treatment effect to score values marginally above the cutoff. \citet{Bertanha-Imbens_2020_JBES} exploit variation in treatment assignment generated by imperfect treatment compliance imposing independence between potential outcomes and compliance types to extrapolate a single-cutoff fuzzy RD treatment effect (i.e., a local average treatment effect at the cutoff) away from the cutoff. Our paper also belongs to this second type, as it relies on within-design information, using only the score and outcome in the Multi-Cutoff RD design.

\citet{Cattaneo-Keele-Titiunik-VazquezBare_2016_JOP} introduced the causal Multi-Cutoff RD framework, which we employ herein, and studied the properties of normalizing-and-pooling estimation and inference in that setting. Building on that paper, \citet{Bertanha_2020_JoE} discusses estimation and inference of an average treatment effect across multi-cutoffs, assuming away cutoff-specific treatment effect heterogeneity. Neither of these papers addressed the topic of RD treatment effect extrapolation across different levels of the score variable, which is the main goal and innovation of the present paper.

All the papers mentioned above focus on extrapolation of RD treatment effects away from the cutoff by relying on continuity-based methods for identification, estimation and inference, which are implemented using local polynomial regression \citep{Fan-Gijbels_1996_Book}. As pointed out by a reviewer, an alternative approach to analyzing RD designs is to employ the local randomization framework introduced by \cite{Cattaneo-Frandsen-Titiunik_2015_JCI}. This framework has been later used in the context of geographic RD designs \citep{Keele-Titiunik-Zubizarreta_2015_JRSSA}, principal stratification \citep{Li-Mattei-Mealli_2015_AoAS}, and kink RD designs \citep{Ganong-Jager_2018_JASA}, among other settings. More recently, this alternative RD framework was expanded to allow for finite-sample falsification testing and for local regression adjustments \citep{Cattaneo-Titiunik-VazquezBare_2017_JPAM}. See also \cite{Sekhon-Titiunik_2016_ObsStud,Sekhon-Titiunik_2017_AIE} for further conceptual discussions, \cite{Cattaneo-Titiunik-VazquezBare_2020_Sage} for another review, and \cite{Cattaneo-Idrobo-Titiunik_2020_Book} for a practical introduction.

Local randomization RD methods implicitly give extrapolation within the neighborhood where local randomization is assumed to hold because they assume a parametric (usually constant) treatment effect model as a function of the score. However, those methods cannot aid in extrapolating RD treatment effects beyond such neighborhood without additional assumptions, which is precisely our goal. Since local randomization methods explicitly view RD designs as local randomized experiments, we can summarize the key conceptual distinction between that literature and our paper as follows: available local randomization methods for RD designs have only internal validity (i.e., within the local randomization neighborhood), while our proposed method seeks to achieve external validity (i.e., outside the local randomization neighborhood), which we achieve by exploiting the presence of multiple cutoffs (akin to multiple local experiments) together with an additional identifying assumption within the continuity-based approach to RD designs (i.e., parallel control regression functions across cutoffs).

Our core extrapolation idea can be developed within the local randomization framework, albeit under considerably stronger assumptions. To conserve space, in the supplemental appendix, we discuss multi-cutoff extrapolation of RD treatment effects using local randomization ideas, and develop randomization-based estimation and inference methods \citep{Rosenbaum_2010_Book,Imbens-Rubin_2015_Book}. We also empirically illustrate these methods in the supplemental appendix.


\section{Extrapolation in Multi-Cutoff RD Designs} \label{sec:methods}

We assume $(Y_i,X_i,C_i,D_i)$, $i=1,2,\dots,n$, is an observed random sample, where $Y_i$ is the outcome of interest, $X_i$ is the score (or running variable), $C_i$ is the cutoff indicator, and $D_i$ is a treatment status indicator. We assume the score has a continuous positive density $f_X(x)$ on the support $\mathcal X$. Unlike the canonical RD design where the cutoff is a fixed scalar, in the Multi-Cutoff RD design the cutoff faced by unit $i$ is the random variable $C_i$ taking values in a set $\mathcal C\subset \mathcal X$. For simplicity, we consider two cutoffs: $\mathcal{C} = \{\L,\H\}$, with $\L<\H$ and $\L,\H \in \mathcal X$. Extensions to more than two cutoffs and to geographic and multi-score RD designs are conceptually straightforward, and hence discussed in the supplemental appendix.

The conditional density of the score at each cutoff is $f_{X|C}(x|c)$, $c\in\mathcal C$. In sharp RD designs treatment assignment and status are identical, and hence $D_i=\mathbbm{1}(X_i\ge C_i)$. Section \ref{sec:extensions} discusses an extension to fuzzy RD designs. Finally, we let $Y_{i}(1)$ and $Y_{i}(0)$ denote the potential outcomes of unit $i$ under treatment and control, respectively, and $Y_i=D_iY_{i}(1)+(1-D_i)Y_{i}(0)$ is the observed outcome.

The potential outcome regression functions are $\mu_{d,c}(x) = \E[Y_{i}(d)|X_i=x,C_i=c]$, for $d=0,1$. We express all parameters of interest in terms of the ``response'' function
\begin{equation}\label{cef}
\tau_c(x) =\E[Y_i(1)-Y_i(0)\mid X_i=x,C_i=c].
\end{equation}
This function measures the treatment effect for the subpopulation exposed to cutoff $c$ when the running variable takes the value $x$. For a fixed cutoff $c$, it records how the treatment effect for the subpopulation exposed to this cutoff varies with the running variable.  As such, it captures a key quantity of interest when extrapolating the RD treatment effect. The usual parameter of interest in the standard (single-cutoff) RD design is a particular case of $\tau_c(x)$ when cutoff and score coincide:
\begin{equation*}\label{eq:RD-single}
\tau_c(c)=\E[Y_i(1)-Y_i(0)\mid X_i=c,C_i=c]=\mu_{1,c}(c)-\mu_{0,c}(c) \text{.}
\end{equation*}
It is well known that, via continuity assumptions, the function $\tau_c(x)$ is nonparametrically identifiable at the single point $x=c$. Our approach exploits the presence of multiple cutoffs to identify this function at other points on a portion of the support of the score variable.

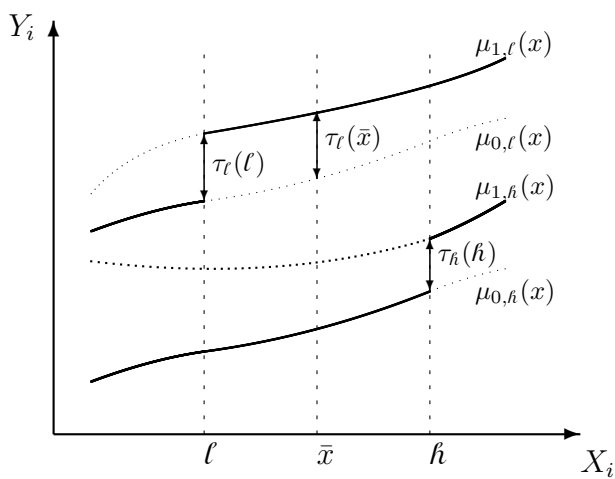
\begin{figure}
	\centering
	\setlength{\unitlength}{1cm}
	\begin{picture}(8,6)(-0.5,-0.5)
	\linethickness{0.25pt}
	
	\thicklines
	\put(0,0){\vector(1,0){7}}
	\put(0,0){\vector(0,1){5.5}}
	\put(7,-0.5){$X_i$}
	\put(-0.6,5.3){$Y_i$}
	
	\thinlines
	\put(2.0,3.5){\colorbox{white}{\footnotesize $\tau_\L(\L)$}}
	\put(2.0,3.1){\vector(0,1){.9}}
	\put(2.0,4.0){\vector(0,-1){.9}}
	
	\put(3.5,3.85){\colorbox{white}{\footnotesize $\tau_\L(\x)$}}
	\put(3.5,3.4){\vector(0,1){.9}}
	\put(3.5,4.2){\vector(0,-1){.8}}
	
	\thinlines
	\put(5.0,2.3){\colorbox{white}{\footnotesize $\tau_\H(\H)$}}
	\put(5.0,1.9){\vector(0,1){.7}}
	\put(5.0,2.6){\vector(0,-1){.7}}
	
	\multiput(2,0)(0,0.2){26}{\line(0,1){0.05}}
	\put(2,-0.4){$\L$}
	\multiput(3.5,0)(0,0.2){26}{\line(0,1){0.05}}
	\put(3.5,-0.4){$\x$}
	\multiput(5,0)(0,0.2){26}{\line(0,1){0.05}}
	\put(5,-0.4){$\H$}
	
	\thinlines \qbezier[ 20](0.5,3.2)(1.0,3.8)(2.0,4.0)
	\thicklines\qbezier[500](2.0,4.0)(5.0,4.5)(6.0,5.0)
	\put(5.6,5.1){\footnotesize\textcolor{black}{$\mu_{1,\L}(x)$}}
	
	\thicklines\qbezier[500](0.5,2.7)(1.3,3.0)(2.0,3.1)
	\thinlines \qbezier[ 30](2.0,3.1)(3.5,3.3)(5.0,3.9)
	\thinlines \qbezier[ 10](5.0,3.9)(5.5,4.1)(6.0,4.2)
	\put(5.6,3.85){\footnotesize\textcolor{black}{$\mu_{0,\L}(x)$}}
	
	\thicklines\qbezier[ 50](0.5,2.3)(3.0,2.0)(5.0,2.6)
	\thicklines\qbezier[500](5.0,2.6)(5.5,2.8)(6.0,3.1)
	\put(5.6,3.2){\footnotesize\textcolor{black}{$\mu_{1,\H}(x)$}}
	
	\thicklines\qbezier[500](0.5,0.7)(1.3,1.0)(2.0,1.1)
	\thicklines\qbezier[500](2.0,1.1)(3.5,1.3)(5.0,1.9)
	\thinlines \qbezier[ 10](5.0,1.9)(5.5,2.1)(6.0,2.2)
	\put(5.6,1.8){\footnotesize\textcolor{black}{$\mu_{0,\H}(x)$}}

	\end{picture}
	\caption{Estimands of interest with two cutoffs.\label{fig:estimands}}
\end{figure}

Figure \ref{fig:estimands} contains a graphical representation of our extrapolation approach for Multi-Cutoff RD designs. In the plot, there are two populations, one exposed to a low cutoff $\L$, and another exposed to a high cutoff $\H$. The RD effects for each subpopulation are, respectively, $\tau_\L(\L)$ and $\tau_\H(\H)$. We seek to learn about the effects of the treatment at points other than the particular cutoff to which units were exposed, such as the point $\bar{x}$ in Figure \ref{fig:estimands}. Below, we develop a framework for the identification of $\tau_\L(x)$ for  $\L<x\leq\H$ so that we can assess what would have been the average treatment effect for the subpopulation exposed to the cutoff $\L$ at score values above $\ell$ (illustrated by the effect $\tau_\L(\bar{x})$ in Figure \ref{fig:estimands} for the intermediate point $X_i=\x$).

In our framework, the multiple cutoffs define different subpopulations. In some cases, the cutoff to which a unit is exposed depends only on characteristics of the units, such as when the cutoffs are cumulative and increase as the score falls in increasingly higher ranges. In other cases, the cutoff depends on external features, such as when different cutoffs are used in different geographic regions or time periods. This means that, in our framework, the cutoff $C_i$ acts as an index for different subpopulation ``types'', capturing both observed and unobserved characteristics of the units. 

Given the subpopulations defined by the cutoff values actually used in the Multi-Cutoff RD design, we consider the effect that the treatment would have had for those subpopulations had the units had a higher score value than observed. This is why, in our notation, the index for the cutoff value is fixed, and the index for the score is allowed to vary and is the main argument of the regression functions. This conveys the idea that the subpopulations are defined by the multiple cutoffs actually employed, and our exercise focuses on studying the treatment effect at different score values for those pre-defined subpopulations. For example, this setting covers RD designs with a common running variable but with cutoffs varying by regions, schools, firms, or some other group-type variable. Our method is not appropriate to extrapolate to populations outside those defined by the Multi-Cutoff RD design.

\subsection{Identification Result}

The main challenge to the identification of extrapolated treatment effects in the single-cutoff (sharp) RD design is the lack of observed control outcomes for score values above the cutoff. In the Multi-Cutoff RD design, we still face this challenge for a given subpopulation, but we have other subpopulations exposed to higher cutoff values that, under some assumptions, can aid in solving the missing data problem and identify average treatment effects. Before turning to the formal derivations, we illustrate the idea graphically. 

\begin{figure}
	\centering
	\setlength{\unitlength}{1cm}
	\begin{picture}(8,6)(-0.5,-0.5)
	\linethickness{0.25pt}
	
	\thicklines
	\put(0,0){\vector(1,0){7}}
	\put(0,0){\vector(0,1){5.5}}
	\put(7,-0.5){$X_i$}
	\put(-0.6,5.3){$Y_i$}
	
	\multiput(2,0)(0,0.2){26}{\line(0,1){0.05}}
	\put(2,-0.4){$\L$}
	\multiput(3.5,0)(0,0.2){26}{\line(0,1){0.05}}
	\put(3.5,-0.4){$\x$}
	\multiput(5,0)(0,0.2){26}{\line(0,1){0.05}}
	\put(5,-0.4){$\H$}
	
	\thinlines \qbezier[ 20](0.5,3.2)(1.0,3.8)(2.0,4.0)
	\thicklines\qbezier[500](2.0,4.0)(5.0,4.5)(6.0,5.0)
	\put(5.6,5.1){\footnotesize\textcolor{black}{$\mu_{1,\L}(x)$}}
	
	\thicklines\qbezier[500](0.5,2.7)(1.3,3.0)(2.0,3.1)
	\thinlines \qbezier[ 30](2.0,3.1)(3.5,3.3)(5.0,3.9)
	\thinlines \qbezier[ 10](5.0,3.9)(5.5,4.1)(6.0,4.2)
	\put(5.6,3.85){\footnotesize\textcolor{black}{$\mu_{0,\L}(x)$}}
	
	\put(3.5,3.85){\colorbox{white}{\footnotesize $\tau_\L(\x)$}}
	\put(3.5,3.4){\vector(0,1){.9}}
	\put(3.5,4.2){\vector(0,-1){.8}}
	
	\thicklines\qbezier[500](0.5,0.7)(1.3,1.0)(2.0,1.1)
	\thicklines\qbezier[500](2.0,1.1)(3.5,1.3)(5.0,1.9)
	\thinlines \qbezier[ 10](5.0,1.9)(5.5,2.1)(6.0,2.2)
	\put(5.6,1.8){\footnotesize\textcolor{black}{$\mu_{0,\H}(x)$}}
	
	\put(2.0,1.1){\line(0,1){2}}
	\put(3.5,1.4){\line(0,1){2}}
	
	\put(3.5,4.27){\circle*{.15}}\put(3.6,4.4){\footnotesize\textcolor{black}{$a$}}
	\put(3.5,3.4){\circle*{.15}}\put(3.6,3.2){\footnotesize\textcolor{black}{$b$}}
	\put(3.5,1.4){\circle*{.15}}\put(3.6,1.2){\footnotesize\textcolor{black}{$c$}}
	\put(2.0,1.1){\circle*{.15}}\put(2.1,1.2){\footnotesize\textcolor{black}{$d$}}
	\put(2.0,3.1){\circle*{.15}}\put(2.1,2.9){\footnotesize\textcolor{black}{$e$}}
	\put(2.0,4.0){\circle*{.15}}\put(2.1,4.1){\footnotesize\textcolor{black}{$f$}}
	
	\put(2.1,2.1){\footnotesize\textcolor{black}{$B(\L)$}}
	\put(3.6,2.4){\footnotesize\textcolor{black}{$B(\x)$}}
	
	\end{picture}
	\caption{RD Extrapolation with Constant Bias ($B(\L)=B(\x)$).\label{fig:constantbias}}
\end{figure}
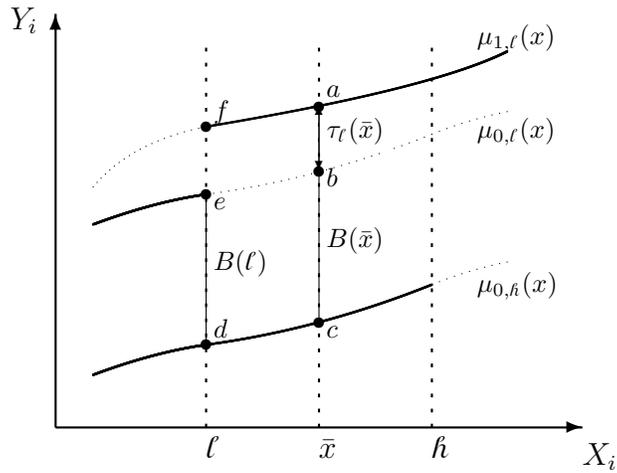

Figure \ref{fig:constantbias} illustrates the regression functions for the populations exposed to cutoffs $\L$ and $\H$, with the function $\mu_{1,\H}(x)$ omitted for simplicity. We seek an estimate of $\tau_\L(\x)$, the average effect of the treatment at the  point $\x \in (\L,\H)$ for the subpopulation exposed to the lower cutoff $\L$. In the figure, this parameter is represented by the segment $\overline{ab}$. The main identification challenge is that we only observe the point $a$, which corresponds to $\mu_{1,\L}(\x)$, the treated regression function for the population exposed to $\L$, but we fail to observe its control counterpart $\mu_{0,\L}(\x)$ (point $b$), because all units exposed to cutoff $\L$ are treated at any $x > \L$. We use the control group of the population exposed to the higher cutoff, $\H$, to infer what would have been the control response at  $\x$ of units exposed to the lower cutoff $\L$. At the point $X_i = \x$, the control response of the population exposed to $\H$ is $\mu_{0,\H}(\x)$, which is represented by the point $c$ in Figure \ref{fig:constantbias}. Since all units in this subpopulation are untreated at $\x$, the point $c$ is identified by the average observed outcomes of the control units in the subpopulation $\H$ at $\x$.

Of course, units facing different cutoffs may differ in both observed and unobserved ways. Thus, there is generally no reason to expect that the average control outcome of the population facing cutoff $\H$ will be a good approximation to the average control outcome of the population facing cutoff $\L$. This is captured in Figure \ref{fig:constantbias} by the fact that $\mu_{0,\L}(\x)\equiv b \neq c \equiv \mu_{0,\H}(\x)$. This difference in untreated potential outcomes for units facing different cutoffs can be interpreted as a bias driven by differences in observed and unobserved characteristics of the different subpopulations, analogous to ``site selection'' bias in multiple randomized experiments. We formalize this idea with the following definition.
\begin{definition}[Cutoff Bias] \label{seldef}
	$B(x,c,c')=\mu_{0,c}(x)-\mu_{0,c'}(x)$, for $c,c'\in \mathcal C$.
	There is bias from exposure to different cutoffs if $B(x,c,c')\ne 0$ for some $c,c'\in \mathcal C$, $c\ne c'$ and for some $x \in \mathcal X$.
\end{definition}

Table \ref{tab:FigParams} defines the parameters associated with the corresponding segments in Figure \ref{fig:constantbias}. The parameter of interest, $\tau_\L(\x)$, is unobservable because we fail to observe $\mu_{0,\L}(\x)$. If we replaced $\mu_{0,\L}(\x)$ with $\mu_{0,\H}(\x)$, we would be able to estimate the distance $\overline{ac}$. This distance, which is observable, is the sum of the parameter of interest,  $\tau_\L(\x)$, plus the bias $B(\x,c,c')$ that arises from using the control group in the $\H$ subpopulation instead of the control group in the $\L$ subpopulation. Graphically, $\overline{ac} = \overline{ab} + \overline{bc}$. Since we focus on the two-cutoff case, we denote the bias by $B(\bar{x})$ to simplify the notation.

We use the distance between the control groups facing the two different cutoffs at a point where both are observable, to approximate the unobservable distance between them at $\x$---that is, to approximate the bias $B(\x)$. As shown in the figure, at $\L$, all units facing cutoff $\H$ are controls and all units facing cutoff $\L$ are treated. But under standard RD assumptions, we can identify $\mu_{0,\L}(\L)$ using the observations in the $\L$ subpopulation whose scores are just below $\L$. Thus, the bias term $B(\L)$, captured in the distance $\overline{ed}$, is estimable from the data.

Graphically, we can identify the extrapolation parameter $\tau_\L(\x)$ assuming that the observed difference between the control functions $\mu_{0,\L}(\cdot)$ and $\mu_{0,\H}(\cdot)$ at $\L$ is constant for all values of the score:
\begin{align*}
\overline{ac} - \overline{ed} & =  \left\{ \mu_{1,\L}(\x) - \mu_{0,\H}(\x)\right\} - \left\{ \mu_{0,\L}(\L) - \mu_{0,\H}(\L)\right\}\\
& = \left\{ \tau_\L(\x) + B(\x)\right\} - \left\{ B(\L) \right\}\\
& = \tau_\L(\x) \text{.}
\end{align*}

We now formalize this intuitive result employing standard continuity assumptions on the relevant regression functions. We make the following assumptions. 

\begin{assumption}[Continuity]\label{assu:continuity}
$\mu_{d,c}(x)$ is continuous in $x\in [\L,\H]$ for $d=0,1$ and for all $c$.
\end{assumption}

The observed outcome regression functions are $\mu_c(x) = \E[Y_i|X_i=x,C_i=c]$, for $c\in\mathcal C= \{\L, \H\}$, and note that by standard RD arguments $\mu_{0,c}(c) = \lim_{\varepsilon \uparrow 0} \mu_{c}(c+\varepsilon)$ and $\mu_{1,c}(c) = \lim_{\varepsilon \downarrow 0} \mu_{c}(c+\varepsilon)$. Furthermore, $\mu_{0,\H}(x) = \mu_{\H}(x)$ and $\mu_{1,\L}(x) = \mu_{\L}(x)$ for all $x \in (\L, \H)$.

Our main extrapolation assumption requires that the bias not be a function of the score, which is analogous to the parallel trends assumption in the difference-in-differences design. 

\begin{assumption}[Constant Bias]\label{assu:constantbias} $B(\L) = B(x)$ for all $x \in (\L, \H)$.
\end{assumption}

While technically our identification result only needs this condition to hold at $x=\x$, in practice it may be hard to argue that the equality between biases holds at a single point. Combining the constant bias assumption with the continuity-based identification of the conditional expectation functions allows us to express the unobservable bias for an interior point, $\x \in (\L,\H)$, as a function of estimable quantities. The bias at the low cutoff $\L$ can be written as 
\begin{align*}
B(\L) &= \lim_{\varepsilon\uparrow 0}\mu_{\L}(\L+\varepsilon) - \mu_{\H}(\L) \text{.}
\end{align*} 
Under Assumption \ref{assu:constantbias}, we have
	\[\mu_{0,\L}(\x) = \mu_{\H}(\x) + B(\L), \qquad \x \in (\L,\H),\] 
that is, the average control response for the $\L$ subpopulation at the interior point $\x$ is equal to the average observed response for the $\H$ subpopulation at the same point, plus the difference in the average control responses between both subpopulations at the low cutoff $\L$.  This leads to our main identification result. 
\begin{theorem}[Extrapolation]\label{thm:extrapol} Under Assumptions \ref{assu:continuity} and \ref{assu:constantbias}, for any point $\x \in (\L,\H )$,
	\[\tau_{\L}(\x) = \mu_{\L}(\x) -  [\mu_{\H}(\x) + B(\L)].\] 
\end{theorem}

This result can be extended to hold for $\bar{x}\in(\L,\H]$ by using side limits appropriately. In Section \ref{sec:paralleltrends}, we discuss two approaches to provide empirical support for the constant bias assumption. We extend our result to  Fuzzy RD designs in Section \ref{sec:extensions}, and allow for non-parallel control regression functions and pre-intervention covariate-adjustment in the supplemental appendix. 

While we develop our core idea for extrapolation from ``left to right'', that is, from a low cutoff to higher values of the score, it follows from the discussion above that the same ideas could be developed for extrapolation from ``right to left''. Mathematically, the problem is symmetric and hence both extrapolations are equally viable. However, conceptually, there is an important asymmetry. Theorem \ref{thm:extrapol} requires the regression functions for \textit{control} units to be parallel over the extrapolation region (Assumption \ref{assu:constantbias}), while a version of this theorem for ``right to left'' extrapolation would require that the regression functions for \textit{treated} units be parallel. These two identifying assumptions are not symmetric because the latter effectively imposes a constant treatment effect assumption across cutoffs (for different values of the score), while the former does not because it pertains to control units only.

\subsection{Estimation and Inference}\label{sec:estimation}

We estimate all (identifiable) conditional expectations $\mu_{d,c}(x)=\E[Y_i(d)|X_i=x,C_i=c]$ using nonparametric local polynomial methods, employing second-generation MSE-optimal bandwidth selectors and robust bias correction inference methods. See \citet{Calonico-Cattaneo-Titiunik_2014_ECMA}, \citet{Calonico-Cattaneo-Farrell_2018_JASA,Calonico-Cattaneo-Farrell_2020_ECTJ,Calonico-Cattaneo-Farrell_2020_wp}, and \cite{Calonico-Cattaneo-Farrell-Titiunik_2019_RESTAT} for more methodological details, and \citet{Calonico-Cattaneo-Farrell-Titiunik_2017_Stata} and \citet{Calonico-Cattaneo-Farrell_2019_JSS} for software implementation. See also \citet{Hyytinen-etal_2018_QE}, \citet{Ganong-Jager_2018_JASA} and \citet{Dong-Lee-Gou_2020_wp} for some recent applications and empirical testing of those methods.

To be more precise, a generic local polynomial estimator is $\hat{\mu}_{d,c}(x)=\mathbf{e}_0'\hat{\boldsymbol{\beta}}_{d,c}(x)$, where
\[\hat{\boldsymbol{\beta}}_{d,c}(x) = \argmin_{\mathbf{b}\in\mathbbm{R}^{p+1}}\sum_{i=1}^n(Y_i-\mathbf{r}_p(X_i-x)'\mathbf{b})^2K\left(\frac{X_i-x}{h}\right)\1(C_i=c)\1(D_i=d),\]
$\mathbf{e}_0$ is a vector with a one in the first position and zeros in the rest, $\mathbf{r}_p(\cdot)$ is a polynomial basis of order $p$, $K(\cdot)$ is a kernel function, and $h$ a bandwidth. For implementation, we set $p=1$ (local-linear), $K$ to be the triangular kernel, $h$ to be a MSE-optimal bandwidth selector, unless otherwise noted. Then, given the two cutoffs $\L$ and $\H$ and an extrapolation point $\bar{x}\in(\L,\H]$, the extrapolated treatment effect at $\bar{x}$ for the subpopulation facing cutoff $\L$ is estimated as
\[\hat{\tau}_\L(\bar{x})
  =\hat{\mu}_{1,\L}(\bar{x})-\hat{\mu}_{0,\H}(\bar{x})-\hat{\mu}_{0,\L}(\L)+\hat{\mu}_{0,\H}(\L).
\]

The estimator $\hat{\tau}_\L(\bar{x})$ is a linear combination of nonparametric local polynomial estimators at boundary and at interior points depending on the choice of $\x$ and data availability. Hence, optimal bandwidth selection and robust bias-corrected inference can be implemented using the methods and software mentioned above. By construction, $\hat{\mu}_{d,\L}(\cdot)$ and $\hat{\mu}_{0,\H}(\cdot)$ are independent because the observations used for estimation come from different subpopulations. Similarly, $\hat{\mu}_{0,\L}(\cdot)$ and $\hat{\mu}_{1,\L}(\cdot)$ are independent since the first term is estimated using control units whereas the second term uses treated units. On the other hand, in finite samples, $\hat{\mu}_{0,\H}(\L)$ and $\hat{\mu}_{0,\H}(\bar{x})$ can be correlated if the bandwidths used for estimation overlap (or, alternatively, if $\L$ and $\bar{x}$ are close enough), in which case we account for such correlation in our inference results. More precisely, $\V[\hat{\tau}_\L(\bar{x})|\mathbf{X}]
 =\V[\hat{\mu}_{1,\L}(\bar{x})|\mathbf{X}]
 +\V[\hat{\mu}_{0,\H}(\bar{x})|\mathbf{X}]
 +\V[\hat{\mu}_{0,\L}(\L)|\mathbf{X}]
 +\V[\hat{\mu}_{0,\H}(\L)|\mathbf{X}]
 -2\cov(\hat{\mu}_{0,\H}(\L),\hat{\mu}_{0,\H}(\bar{x})|\mathbf{X})$,
where $\mathbf{X}=(X_1,X_2,\dots,X_n)'$.

Precise regularity conditions for large sample validity of our estimation and inference methods can be found in the references given above. The replication files contain details on practical implementation. 

\subsection{Assessing the Validity of the Identifying Assumption}\label{sec:paralleltrends}

Assessing the validity of our extrapolation strategy should be a key component of empirical work using these methods. In general, while the assumption of constant bias is not testable, this assumption can be tested indirectly via falsification. While a falsification test cannot demonstrate that an assumption holds, it can provide persuasive evidence that an assumption is implausible. We now discuss two strategies for falsification tests to probe the credibility of the constant bias assumption that is at the center of our extrapolation approach. 

The first falsification approach relies on a global polynomial regression. We test globally whether the conditional expectation functions of the two control groups are parallel below the lowest cutoff. One way to implement this idea, given the two cutoff points $\L < \H$, is to test $\boldsymbol{\delta}=\mathbf{0}$ based on the regression model
\[Y_i = \alpha + \beta\1(C_i=\H) + \mathbf{r}_p(X_i)'\boldsymbol{\gamma}
      + \1(C_i=\H)\mathbf{r}_p(X_i)'\boldsymbol{\delta} + u_i, \qquad \E[u_i|X_i,C_i]=0,
\]
only for units with $X_i<\L$. In words, we employ a $p$-th order global polynomial model to estimate the two regression functions $\E[Y_i|X_i=x,X_i<\L,C_i=\L]$ and $\E[Y_i|X_i=x,X_i<\L,C_i=\H]$, separately, and construct a hypothesis test for whether they are equal up to a vertical shift (i.e., the null hypothesis is $\mathsf{H}_0:\boldsymbol{\delta}=\mathbf{0}$). This approach is valid under standard regularity conditions for parametric least squares regression. This approach could also be justified from a nonparametric series approximation perspective, under additional regularity conditions.

The second falsification approach employs nonparametric local polynomial methods. We test for equality of the derivatives of the conditional expectation functions for values $x<\L$. Specifically, we test for $\mu^{(1)}_{\L}(x)=\mu^{(1)}_\H(x)$ for all $x<\L$, where $\mu^{(1)}_{\L}(x)$ and $\mu^{(1)}_{\H}(x)$ denote the derivatives of $\E[Y_i|X_i=x,X_i<\L,C_i=\L]$ and $\E[Y_i|X_i=x,X_i<\L,C_i=\H]$, respectively. This test can be implemented using several evaluation points, or using a summary statistic such as the supremum. Validity of this approach is also justified using nonparametric estimation and inference results in the literature, under regularity conditions.

\section{Extrapolating the Effect of Loan Access on College Enrollment}\label{sec:analysis}

We use our proposed methods to investigate the external validity of the ACCES program RD effects. As mentioned above, our sample has observations exposed to two cutoffs, $\L = -850$ and $\H = -571$. We begin by extrapolating the effect to the point $\x = -650$; our focus is thus the effect of eligibility for ACCES on whether the student enrolls in a higher education program for the subpopulation exposed to cutoff 850 when their SABER 11 score is $650$.

As described in Section \ref{sec:multicutoffs}, the observations facing cutoff $\L$ correspond to years 2000 to 2008, whereas observations facing cutoff $\H$ correspond to years 2009 and 2010. Our identification assumption allows these two groups to differ in observable and unobservable characteristics so long as the difference between the conditional expectations of their control potential outcomes is constant as a function of the running variable. In addition, our approach relies on the assumption that the underlying population does not change over time (which is implicit in our notation). We offer empirical support for these assumptions in two ways. First, we implement the tests discussed in Section \ref{sec:paralleltrends} to assess the plausibility of Assumption \ref{assu:constantbias}. In addition, Section SA-2 in the Supplemental Appendix shows that our results remain qualitatively unchanged when restricting the empirical analysis to the period 2007-2010, which reduces the (potential) heterogeneity of the underlying populations over time. 

\begin{figure}
	\centering
	\caption{Parallel trends test}
	\begin{subfigure}[b]{0.45\textwidth}
		\includegraphics[width=\textwidth]{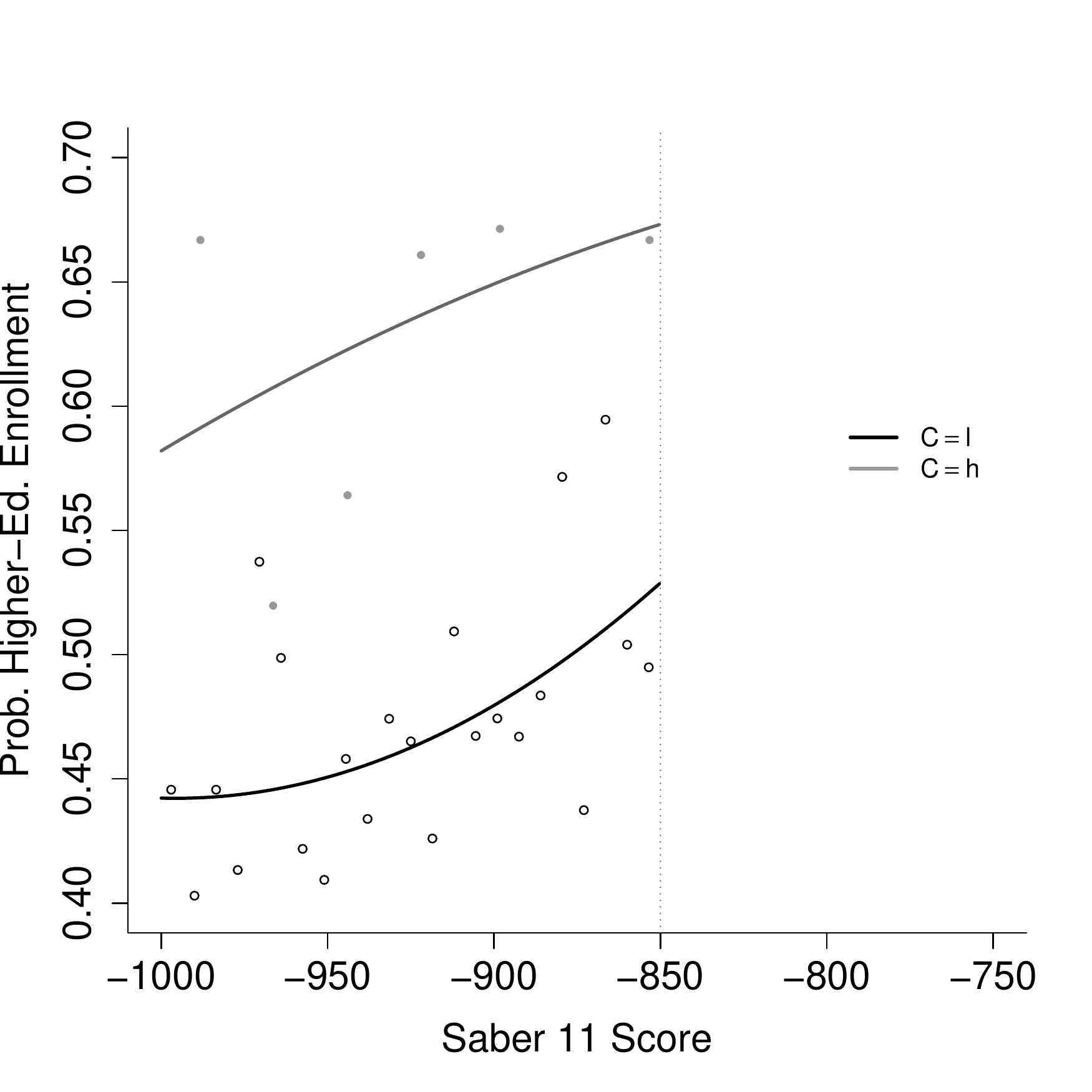}
		\caption{Global Polynomial Approach \label{fig:rdmcplot_trends}}
		\vspace{-.25in}
	\end{subfigure}
	\begin{subfigure}[b]{0.45\textwidth}
		\includegraphics[width=\textwidth]{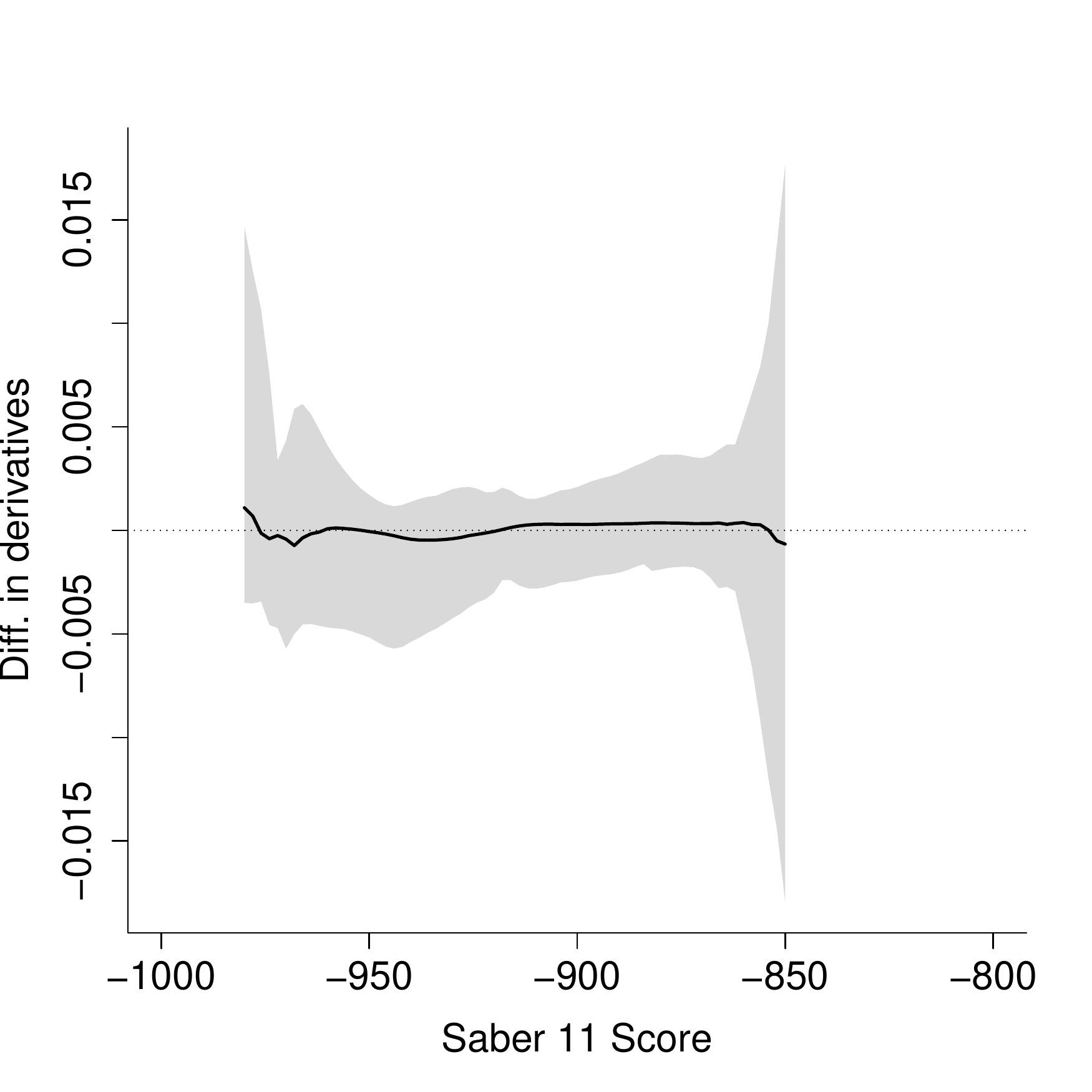}
		\caption{Local Polynomial Approach \label{fig:rdmcplot_trends_deriv}}
		\vspace{-.25in}
	\end{subfigure}
	
	\begin{flushleft}\footnotesize Notes. Panel (a) plots regression functions estimated using a quadratic global polynomial regression. Panel (b) plots the difference in derivatives at several points, estimated using a local quadratic polynomial regression (solid line). The gray area represents the RBC 95\% (pointwise) confidence intervals.\end{flushleft}
\end{figure}

We begin by assessing the validity of our constant bias assumption with the methods described in Section \ref{sec:paralleltrends}. The results can be seen in Tables \ref{tab:trendsgl} and \ref{tab:trendsloc}. Specifically, Table \ref{tab:trendsgl} reports results employing global polynomial regression, which does not reject the null hypothesis of parallel trends. Figure \ref{fig:rdmcplot_trends} offers a graphical illustration. Table \ref{tab:trendsloc} shows the results for the local polynomial approach, which again does not reject the null hypothesis. Additionally, Figure \ref{fig:rdmcplot_trends_deriv} plots the difference in derivatives (solid line) between groups estimated nonparametrically at ten evaluation points below $\L$, along with pointwise robust bias-corrected confidence intervals (dashed lines). The figure reveals that the difference in derivatives is not significantly different from zero.

\begin{sidewaysfigure}
	\begin{center}
		\caption{RD and Extrapolation Effects of ACCES loan eligibility on Higher Education Enrollment\label{fig:RD-results}}
		\begin{subfigure}[b]{0.32\textwidth}
			\includegraphics[width=1\textwidth]{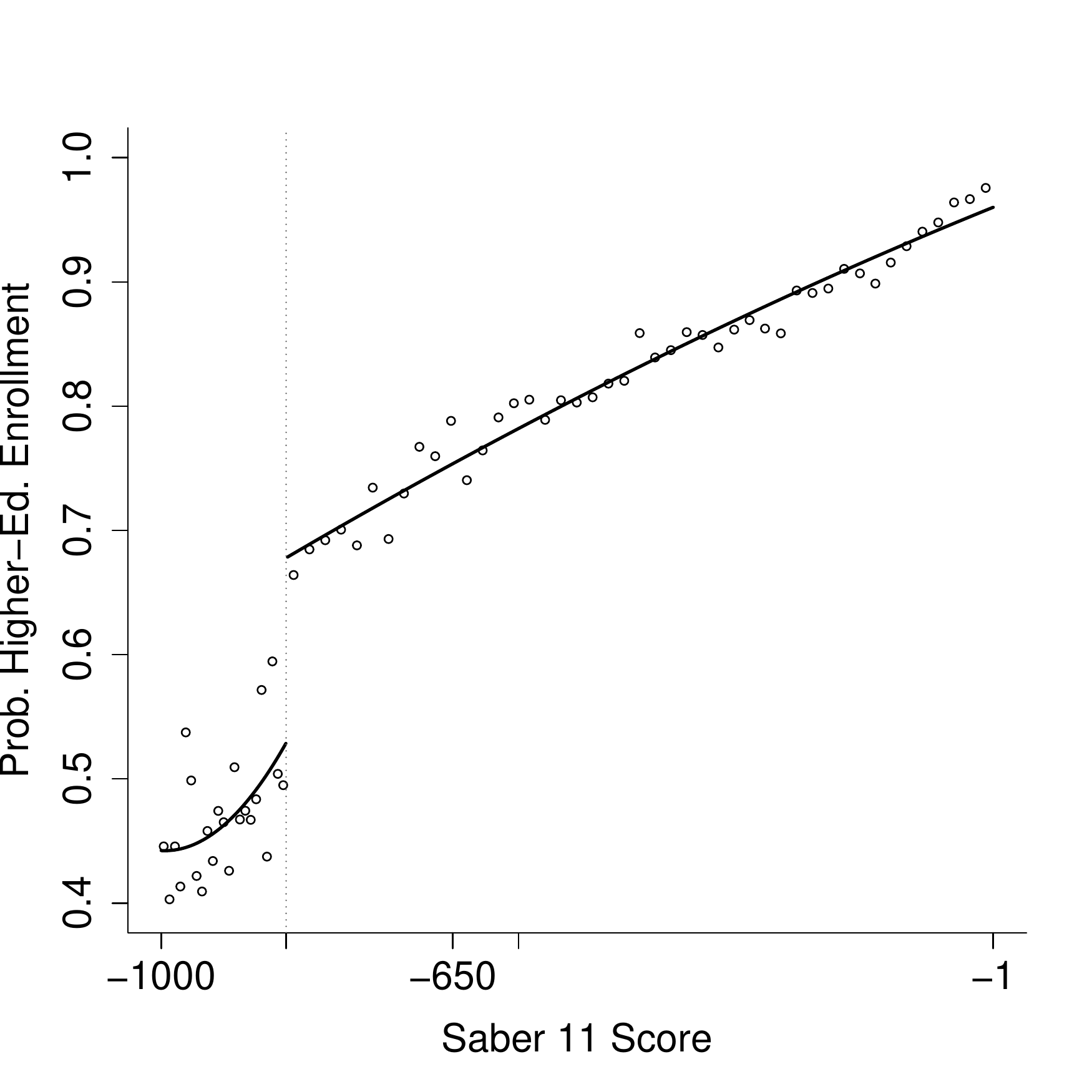}
			\caption{Effect at Cutoff 850}
			\label{subfig:lowC}
		\end{subfigure} 
		\begin{subfigure}[b]{0.32\textwidth}
			\includegraphics[width=1\textwidth]{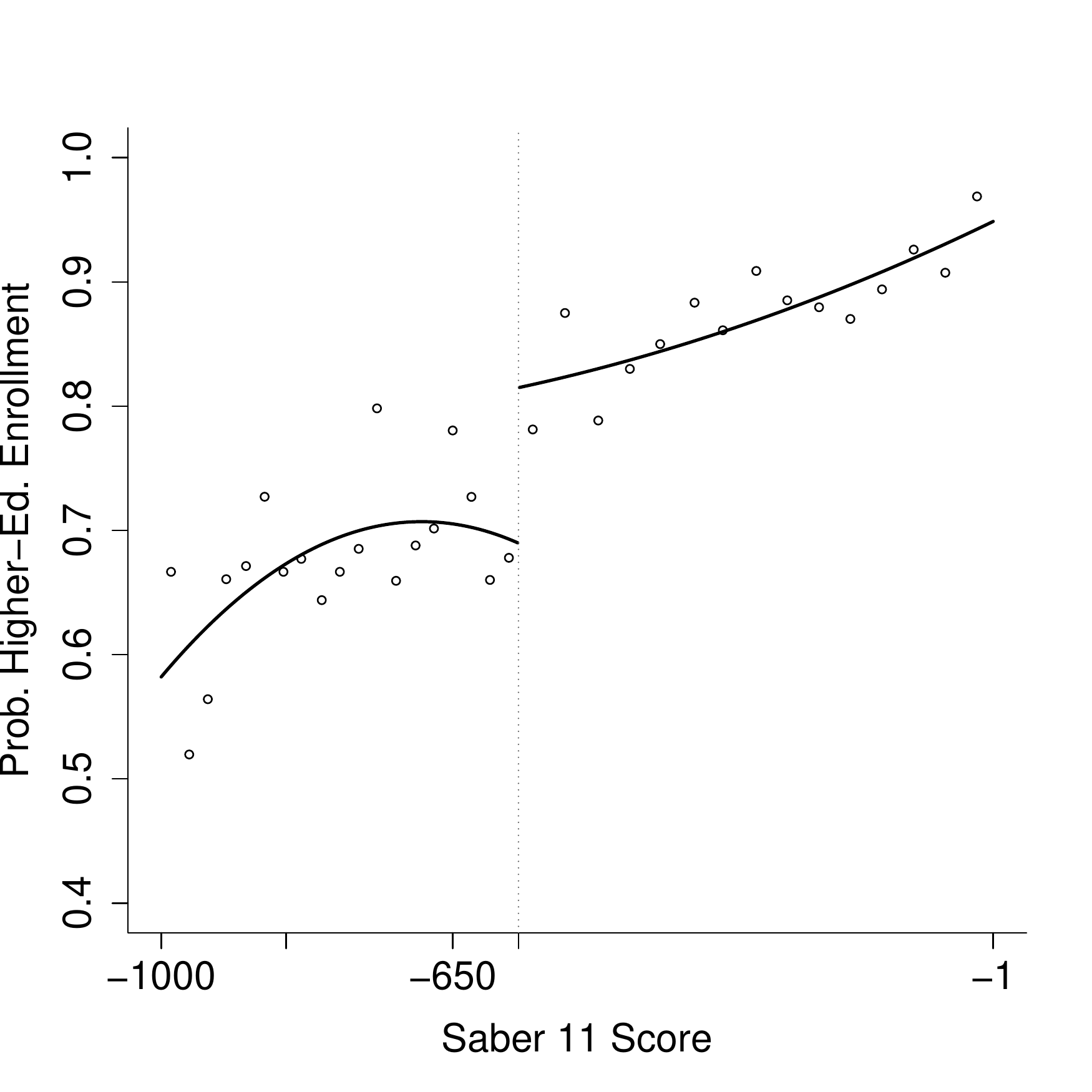}
			\caption{Effect at Cutoff 571}
			\label{subfig:highC}
		\end{subfigure}	
		\begin{subfigure}[b]{0.32\textwidth}
			\includegraphics[width=1\textwidth]{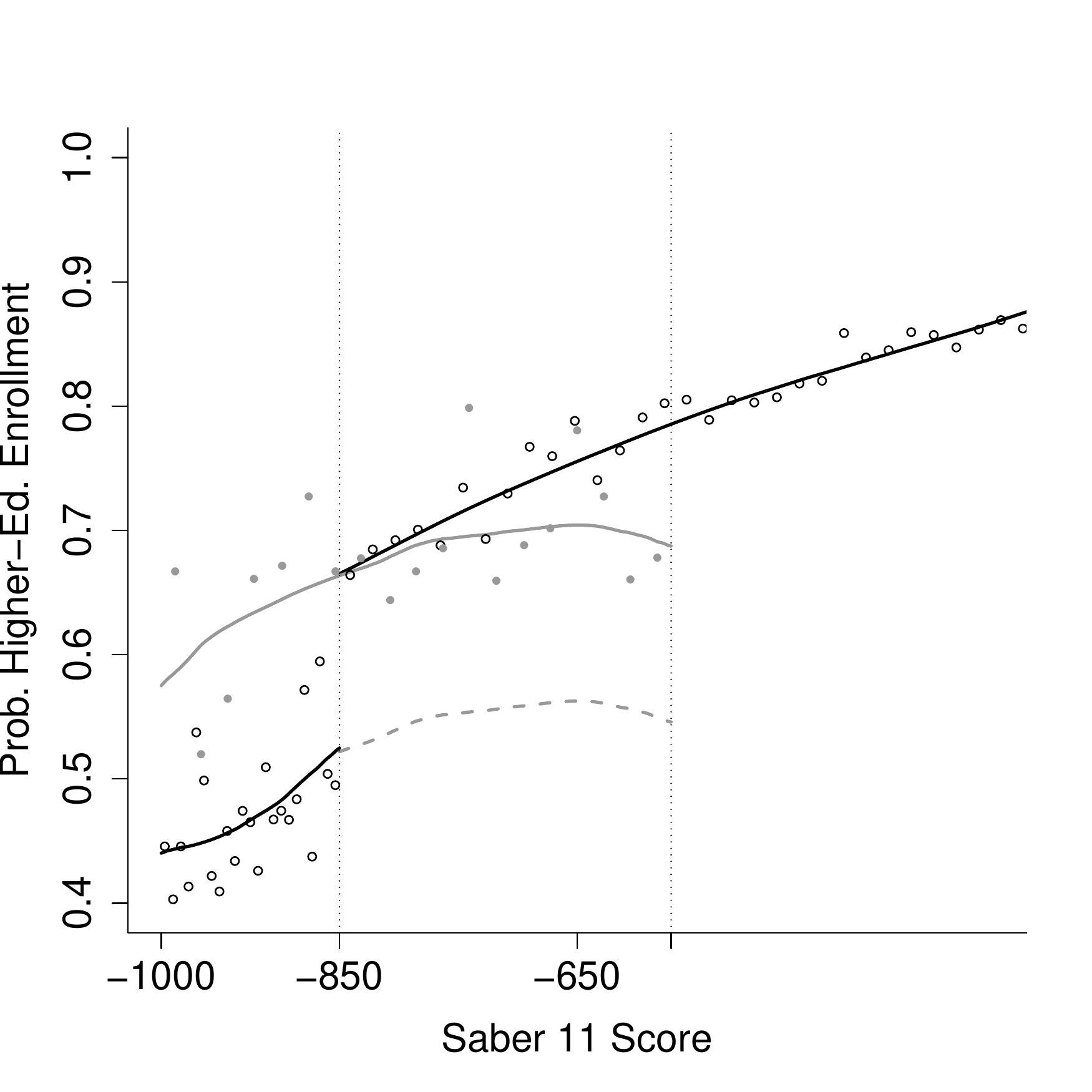}
			\caption{Extrapolated Effect at Cutoff 650}
			\label{subfig:extrapol}
		\end{subfigure}
	\end{center}
	Notes. Panels (a) and (b) show the RD plots for the cutoff-specific effects at the low and high cutoff, respectively. Panel (c) shows the nonparametric local-polynomial estimates of the regression functions for the low-cutoff (solid black line) and high-cutoff (gray line) groups. The dashed line represents the nonparametric local-polynomial estimate of the imputed regression function for control units facing the low cutoff. 
\end{sidewaysfigure}

As discussed in Section \ref{sec:pooled} and Table \ref{tab:results}, the pooled RD estimated effect is 0.125 with a RBC confidence interval of $[0.012,0.219]$. The single-cutoff effect at $-850$ is $0.137$ with $95\%$ RBC confidence interval of $[0.036,0.232]$, and the effect at $-571$ is somewhat higher at $0.169$, with $95\%$ RBC confidence interval of $[-0.039 , 0.428]$. These estimates based on single-cutoffs are illustrated in Figures \ref{fig:RD-results}(\subref{subfig:lowC}) and \ref{fig:RD-results}(\subref{subfig:highC}), respectively.

In finite samples, the pooled estimate may not be a weighted average of the cutoff-specific estimates as it contains an additional term that depends on the bandwidth used for estimation and small sample discrepancies between the estimated slopes for each group. This is evident in Table \ref{tab:results}, where the pooled estimate does not lie between the cutoff specific estimates. This additional term vanishes as the sample size grows and the bandwidths converge to zero, yielding the result in \citet{Cattaneo-Keele-Titiunik-VazquezBare_2016_JOP}. To provide further evidence on the overall effect of the program, we also estimated a weighted average of cutoff-specific effects using estimated weights. This average effect equals 0.156 with a RBC confidence interval of $[0.025,0.314]$. Since this estimate is a proper weighted average of cutoff-specific effects, it may give a more accurate assessment of the overall effect of the program.

The extrapolation results are illustrated in Figure \ref{fig:RD-results}(\subref{subfig:extrapol}) and reported in the last two panels of Table \ref{tab:results}. At the $-650$ cutoff, the treated population exposed to cutoff $-850$ has an enrollment rate of $0.756$, while the control population exposed to cutoff $-571$ has a rate of $0.706$. This naive comparison, however, is likely biased due to unobservable differences between both subpopulations. The bias, which is estimated at the low cutoff $-850$, is $-0.142$, showing that the control population exposed to the $-850$ cutoff has lower enrollment rates at that point than the population exposed to the high cutoff $-571$ (0.525 versus 0.667). The extrapolated effect in the last row corrects the naive comparison according to Theorem \ref{thm:extrapol}. The resulting extrapolated effect is $0.756 - (0.706 - 0.142) = 0.191$ with RBC confidence interval of $[ 0.080, 0.336]$.

\begin{figure}
	\centering
	\caption{Extrapolation treatment effects}
	\begin{subfigure}[b]{0.49\textwidth}
		\includegraphics[scale=.5]{\pathin/rdmcplot_extrapol.pdf}
		\caption{Estimated regression functions.\label{fig:extrapol}}
	\end{subfigure}	
	\begin{subfigure}[b]{0.49\textwidth}
		\includegraphics[scale=.5]{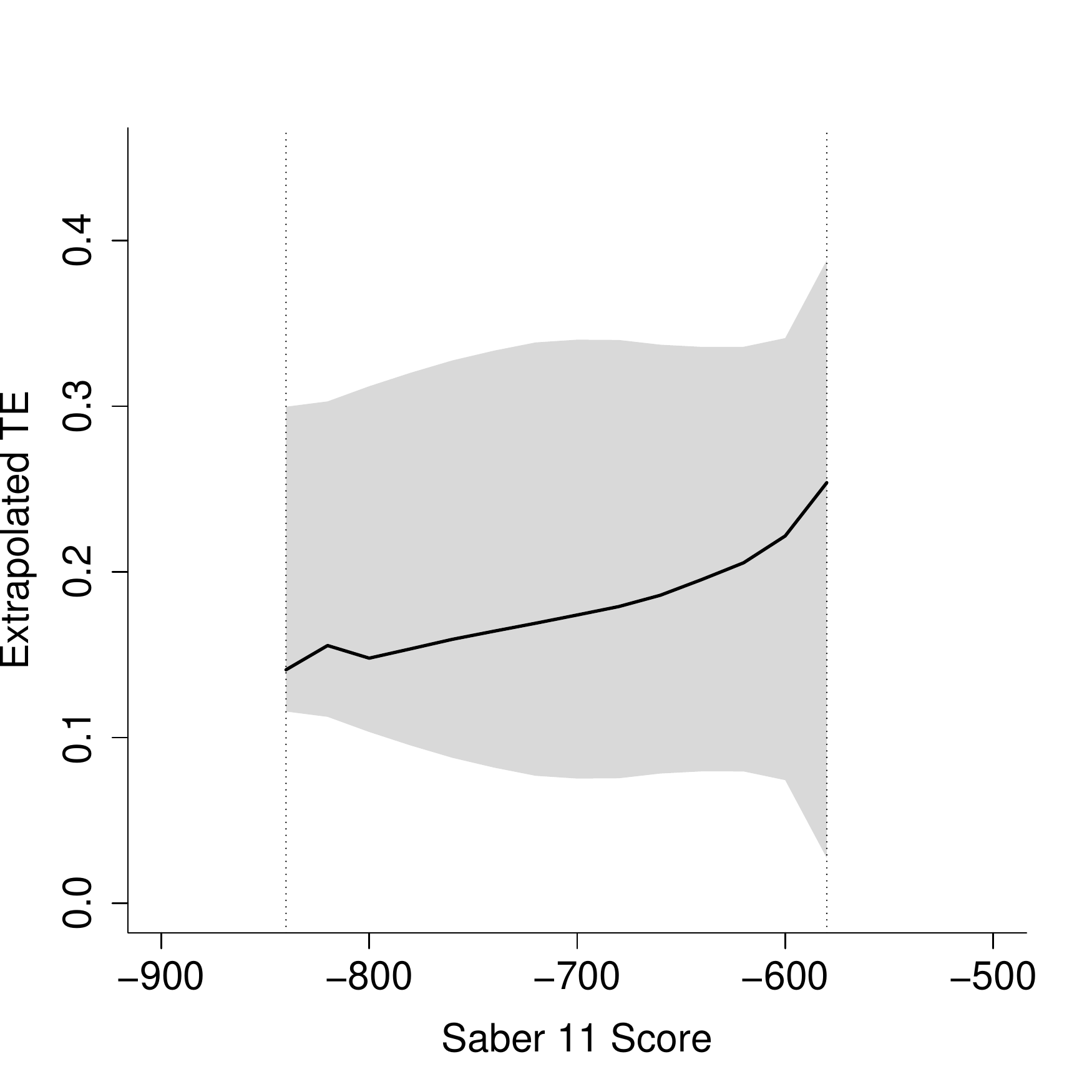}
		\caption{Extrapolation at multiple points.\label{fig:teffects}}
	\end{subfigure}
	
	\begin{flushleft}\footnotesize Notes. Panel (a) shows local-linear estimates of the regression functions using an IMSE-optimal bandwidth for the control and treated groups facing cutoff $\L$ (black solid lines) and for the control group facing cutoff $\H$ (solid gray line). The dashed line represents the extrapolated regression function for the control group facing cutoff $\L$. Panel (b) shows local-linear extrapolation treatment effects estimates at 14 equidistant evaluation points between $-840$ and $-580$. The gray area represents the RBC 95\% (pointwise) confidence intervals. \end{flushleft}
\end{figure}

The choice of the point $-650$ is simply for illustration purposes, and indeed considering a set of evaluation points for extrapolation can give a much more complete picture of the impact of the program away from the cutoff point. In Figures \ref{fig:extrapol} and \ref{fig:teffects}, we conduct this analysis by estimating the extrapolated effect at 14 equidistant points between $-840$ and $-580$. The effects are statistically significant, ranging from around $0.14$ to $0.25$. 


\section{Simulations}\label{sec:simuls}

We report results from a simulation study aimed to assess the performance of the local polynomial methods described in Section \ref{sec:estimation}. We construct $\mu_{0,\H}(x)$ as a fourth-order polynomial where the coefficients are calibrated using the data from our empirical application, and $\mu_{0,\ell}(x) = \mu_{0,\H}(x) + \Delta$. Based on our empirical findings, we set $\Delta=-0.14$ and an extrapolated treatment effect of $\tau_\ell(\bar{x})=0.19$. We consider three sample sizes: $N=1,000$ (``small N''), $N=2,000$ (``moderate N''), and $N=5,000$ (``large N''). To assess the effect of unbalanced sample sizes across evaluation points/cutoffs, our simulation model ensures that some evaluation points/cutoffs have fewer observations than others. In particular, the available sample size to estimate $\mu_\ell(\ell)$ is always less than a third of the sample size available to estimate $\mu_\H(\bar{x})$. We provide all details in the supplemental appendix to conserve space.

The results are shown in Table \ref{tab:simul}. The robust bias-corrected $95\%$ confidence interval for $\tau_\ell(\bar{x})$ has an empirical coverage rate of around 91 percent in the ``small N'' case. This is because one of the parameters, $\mu_\ell(\ell)$, is estimated using very few observations. The empirical coverage rate increases slightly to 92 percent in the ``moderate N'' case, and to about 94 percent in the ``large N'' case. In sum, in our Monte Carlo experiment, we find that local polynomial methods can yield estimators with little bias and RBC confidence intervals with accurate coverage rates for RD extrapolation.


\section{Extension to Fuzzy RD Designs} \label{sec:extensions}

The main idea underlying our extrapolation methods can be extended in several directions that may be useful in other applications. We briefly discuss an extension to Fuzzy RD designs employing a continuity-based approach. In the supplemental appendix we discuss other extensions: covariate adjustments (i.e., ignorable cutoff bias), score adjustments (i.e., polynomial-in-score cutoff bias), many multiple cutoffs, and multiple scores and geographic RD designs.

In the Fuzzy RD design, treatment compliance is imperfect, which is common in empirical applications. For simplicity, we focus on the case of one-sided (treatment) non-compliance: units assigned to the control group comply with their assignment but units assigned to treatment status may not. This case is relevant for a wide array of empirical applications in which program administrators are able to successfully exclude units from the treatment, but cannot force units to actually comply with it.

We employ the Fuzzy Multi-Cutoff RD framework of \citet{Cattaneo-Keele-Titiunik-VazquezBare_2016_JOP}, which builds on the canonical framework of \citet{AngristImbensRubin1996-JASA}. Let $D_i(x,c)$ be the binary treatment indicator and $\underline{x}\leq\bar{x}$. We define compliers as units with $D_i(\underline{x},c)<D_i(\bar{x},c)$, always-takers as units with $D_i(\underline{x},c)=D_i(\bar{x},c)=1$, never-takers as units with $D_i(\underline{x},c)=D_i(\bar{x},c)=0$, and defiers as units with $D_i(\underline{x},c)>D_i(\bar{x},c)$. We assume the following conditions:

\begin{assumption}[Fuzzy RD Design]\label{assu:fuzzy}$ $
\begin{enumerate}
\item Continuity: $\E[Y_i(0)|X_i=x,C_i=c]$ and $\E[(Y_i(1)-Y_i(0))D_i(x,c)|X_i=x,C_i=c]$ are continuous in $x$ for all $c$.
\item Constant bias: $B(\L) = B(x)$ for all $x \in (\L, \H)$.
\item Monotonicity: $D_i(\underline{x},c) \leq D_i(\bar{x},c)$ for all $i$ and for all $\underline{x}\leq\bar{x}$.
\item One-sided noncompliance: $D_i(x,c)=0$ for all $x<c$.
\end{enumerate}
\end{assumption}

The conditions are standard in the fuzzy RD literature and used to identify the local average treatment effect (LATE), which is the treatment effect for units that comply with the RD assignment. The following result shows how to recover a LATE-type extrapolation parameter in this fuzzy RD setting.

\begin{theorem}\label{thm:fuzzy}
Under Assumption \ref{assu:fuzzy},
\[\frac{\mu_{\L}(\x) -  [\mu_{\H}(\x) + B(\L)]}{\E[D_i|X_i=x,C_i=\L]}=\E[Y_i(1)-Y_i(0)|X_i=x,C_i=\L,D_i(x,\L)=1].\]
\end{theorem}

The left-hand side can be interpreted as an ``adjusted'' Wald estimand, where the adjustment allows for extrapolation away from the cutoff point $\L$. More precisely, this theorem shows that under one-sided (treatment) noncompliance we can recover the average extrapolated effect on compliers by dividing the adjusted intention-to-treat parameter by the proportion of compliers.


\section{Conclusion} \label{sec:conclusion}

We introduced a new framework for the extrapolation of RD treatment effects when the RD design has multiple cutoffs. Our approach relies on the assumption that the average outcome difference between control groups exposed to different cutoffs is constant over a chosen extrapolation region. Our method does not require any information external to the design, and can be used whenever two or more cutoffs are used to assign the treatment for different subpopulations, which is a very common feature in many RD applications. Our main extrapolation idea can also be used in settings with more than two cutoffs, multi-scores RD designs \citep{PapayWillettMurnane2011-JoE,ReardonRobinson2012-JREE}, and geographic RD designs \citep{Keele-Titiunik_2015_PA}. In addition, our main idea can be extended to the RD local randomization framework introduced by \cite{Cattaneo-Frandsen-Titiunik_2015_JCI} and \cite{Cattaneo-Titiunik-VazquezBare_2017_JPAM}. These additional results are reported in the supplemental appendix for brevity.

\onehalfspacing

\bibliographystyle{jasa}
\bibliography{Cattaneo-Keele-Titiunik-VazquezBare_2020_JASA--bib}

\clearpage

\begin{table}
	\begin{center}
		\caption{Main Empirical Results for ACCES loan eligibility on Post-Education Enrollment\label{tab:results}}
		\resizebox{.8\textwidth}{!}{\input{\pathin/table_results_650.txt}}
	\end{center}
	\footnotesize Notes. Local polynomial regression estimation with MSE-optimal bandwidth selectors and robust bias corrected inference. See \cite{Calonico-Cattaneo-Titiunik_2014_ECMA} and \cite{Calonico-Cattaneo-Farrell_2018_JASA} for methodological details, and \cite{Calonico-Cattaneo-Farrell-Titiunik_2017_Stata} and \cite{Cattaneo-Titiunik-VazquezBare_2020_Stata} for implementation. ``Eff. $N$'' indicates the effective sample size, that is, the sample size within the MSE-optimal bandwidth. ``Bw'' indicates the MSE-optimal bandwidth.
\end{table}

\begin{table}[h]
	\caption{Segments and Corresponding Parameters in Figure \ref{fig:estimands}}
	\label{tab:FigParams}
	\centering
	\resizebox{1\textwidth}{!}{
		\begin{tabular}{ccl}
			\toprule
			\textbf{Segment} & \textbf{Parameter} & \textbf{Description}\\  
			\midrule 
			$\overline{ab}$ & $\tau_\L(\x) = \underbrace{\mu_{1,\L}(\x)}_\text{Observable}  - \underbrace{\mu_{0,\L}(\x)}_\text{Unobservable}$
			& Extrapolation parameter of interest\vspace{0.2in} \\
			$\overline{bc}$ & $B(\x) = \underbrace{\mu_{0,\L}(\x)}_\text{Unobservable} - \underbrace{\mu_{0,\H}(\x)}_\text{Observable}$
			& Control facing $\L$ vs. control facing $\H$, at $X_i=\x$ \vspace{0.2in} \\
			$\overline{ac}$ & $\tau_\L(\x) + B(\x) = \underbrace{\mu_{1,\L}(\x)}_\text{Observable} - \underbrace{\mu_{0,\H}(\x)}_\text{Observable}$
			& Treated facing $\L$ vs. control facing $\H$, at $X_i=\x$ \vspace{0.2in} \\
			$\overline{ed}$ & $B(\L) = \underbrace{\mu_{0,\L}(\L)}_\text{Observable} - \underbrace{\mu_{0,\H}(\L)}_\text{Observable}$
			& Control facing $\L$ vs. control facing $\H$, at $X_i=\L$ \vspace{0.05in}\\
			\bottomrule
		\end{tabular}
	}
\end{table}

\begin{table}[h]
	\begin{center}
		\caption{Parallel Trends Test: Global Polynomial Approach\label{tab:trendsgl}}
		\resizebox{.5\textwidth}{!}{\input{\pathin/table_results_trendsgl.txt}}
	\end{center}
	\footnotesize Notes. Global (quadratic) polynomial regression with interactions to test for parallel trends between control regression functions for low ($C=\L$) and high ($C=\H$) cutoffs. Estimation and inference is conducted using standard parametric linear least squares methods. $F$-test refers to a joint significance test that the coefficients associated with $\1(C=\H)$, $\1(C=\H)\times Score$ and $\1(C=\H)\times Score^2$ are simultaneously equal to zero.
\end{table}

\begin{table}[h]
	\begin{center}
		\caption{Parallel Trends Test: Local Polynomial Approach\label{tab:trendsloc}}
		\resizebox{.8\textwidth}{!}{\input{\pathin/table_results_trendsloc.txt}}
	\end{center}
	\footnotesize  Notes. Local polynomial methods for testing equality of first derivatives of control regression functions for low ($C=\L$) and high ($C=\H$) cutoffs, over a grid of points below the low ($C=\L$) cutoff. Estimation and robust bias corrected inference is conducted using methods in \cite{Calonico-Cattaneo-Farrell_2018_JASA,Calonico-Cattaneo-Farrell_2020_wp}, implemented via the general purpose software described in \cite{Calonico-Cattaneo-Farrell_2019_JSS}.
\end{table}

\begin{table}
	\begin{center}
		\caption{Simulation Results \label{tab:simul}}
		\resizebox{.8\textwidth}{!}{\input{\pathin/table_results_simul.txt}}
	\end{center}
	\footnotesize Notes. Local polynomial regression estimation with MSE-optimal bandwidth selectors and robust bias corrected inference. See \cite{Calonico-Cattaneo-Titiunik_2014_ECMA} and \cite{Calonico-Cattaneo-Farrell_2018_JASA} for methodological details, and \cite{Calonico-Cattaneo-Farrell-Titiunik_2017_Stata} and \cite{Cattaneo-Titiunik-VazquezBare_2020_Stata} for implementation. ``Eff. $N$'' indicates the effective sample size, that is, the sample size within the MSE-optimal bandwidth. Results from 10,000 simulations.
\end{table}

\clearpage

\end{document}


\title{Extrapolating Treatment Effects in Multi-Cutoff Regression Discontinuity Designs \\ Supplemental Appendix}

\author{
	Matias D. Cattaneo\footnote{Department of Operations Research and Financial Engineering, Princeton University.}\and
	Luke Keele\footnote{Department of Surgery and Biostatistics, University of Pennsylvania.}\and
	Roc\'{i}o Titiunik\footnote{Department of Politics, Princeton University.} \and
	Gonzalo Vazquez-Bare\footnote{Department of Economics, University of California at Santa Barbara.}
}
\maketitle
\setcounter{page}{0}\thispagestyle{empty}

\begin{abstract}
This supplemental appendix includes a description of RD empirical papers with multiple cutoffs, further extensions and generalizations of our main methodological results, and omitted proofs and derivations. In addition, we outline an extension of our methods to the RD local randomization framework introduced by \cite{Cattaneo-Frandsen-Titiunik_2015_JCI} and \cite{Cattaneo-Titiunik-VazquezBare_2017_JPAM}. Finally, \texttt{R} and \texttt{Stata} replication files are available at \url{https://sites.google.com/site/rdpackages/replication/}.\bigskip
\end{abstract}

\newpage

\tableofcontents

\doublespacing\newpage

\section{Empirical Examples of Multi-Cutoff RD Designs}

Table \ref{tab:litreview} collects over $30$ empirical papers in social, behavioral and biomedical sciences, where RD designs with multiple cutoffs are present. In most of these papers, the multi-cutoff feature of the RD designs was not exploited, but rather a pooling-and-normalizing approach was taken to conduct the empirical analysis \citep[see][for methodological background]{Cattaneo-Keele-Titiunik-VazquezBare_2016_JOP}.

\begin{table}[htbp]
	\footnotesize{
		\begin{center}
			\caption{Empirical Papers Employing RD Designs with Multiple Cutoffs.} 
			\label{tab:litreview}
			\resizebox{\textwidth}{!}{\begin{tabular}{lllll}
				\toprule
				\textbf{Citation} & \textbf{Place} & \textbf{Running Variable} & \textbf{Outcome Variable} & \textbf{Cutoffs}\\
				\midrule
				\citet{abdulkadroglu2017regression}	&	US	&	Test scores	&	Test scores	& Many \\
				\citet*{Angrist:1999}	&	Israel	&	Cohort size	&	Test scores	& 2\\
				\citet*{Behaghel:2008} &	France	&	Age	&	Layoff rates & Many \\
				\citet*{Berk:1999} 	&	U.S.	&	Prison score	&	Re-conviction & 4\\
				\citet*{Black:2007}	&	U.S.	&	Training eligibility score	&	Job training and aid & Many\\
				\citet*{Brollo:2012}	&	Brazil	&	Population	&	Federal transfers & Many	\\
				\citet*{Buddelmeyer:2004}	&	Mexico	&	Poverty score	&	Education attainment & 	Many\\
				\citet*{Canton:2004}	&	Mexico	&	Elgibility score	&	College outcomes	& Many\\
				\citet*{Card:2004}	&	U.S.	&	Child age and income	&	Doctor visits	& 2 \\
				\citet*{card2016can}	&	US	&	Elgibility score	&	Test score	& Many\\
				\citet*{Chay:2005} &	Chile	&	Elgibility score	&	School aid	& 13 \\
				\citet*{Chen:2004}	&	U.S.	&	Prison score	&	Rearrest	& 5\\
				\citet*{Chen:2008} &	U.S.	 &	Age	&	Disability awards	& 3 \\
				\citet*{Clark:2009}	&	UK	&	Majority vote	&	Test scores	&  Many\\
				\citet*{crost2014aid} & Phillipines & Poverty index & Conflict & 22 \\
				\citet*{dell2017nation}	&	Vietnam	&	Geographic regions	&	Military strategy by location & Many	\\
				\citet*{Dobkin:2010}	&	U.S.	&	Birthday	&	Education attainment	& 3 \\
				\citet*{Edmonds:2004}	&	S. Africa	&	Age	&	Child outcomes	& 2 - 3\\
                \citet{garibaldi2012college}	&	Italy	&	Income	&	Graduation	& 12 \\
				\citet*{Goodman:2008}	&	U.S.	&	Test score	&	Scholarship	& Many \\
				\citet*{Hjalmarsson:2009} &	U.S.	&	Adjudication score	&	Re-conviction	& 2\\
				\citet*{hoekstra2009effect}	&	US	&	Test and grades	&	Earnings	& Many\\
				\citet*{Hoxby:2000}	&	U.S.	&	Cohort size	&	Test scores	& Many\\
				\citet*{Kane:2003}	&	U.S.	&	GPA	&	College attendance	& Many\\
				\citet*{klavsnja2017incumbency} & Brazil & Margin of victory & Incumbency advantage & Many\\
                \citet*{kirkeboen2016field}	&	Norway	&	Test scores	&	Earnings	& Many\\
				\citet*{Litschig:2013}	&	Brazil	&	Population	&	Poverty reduction	& 17\\
				\citet*{spenkuch2018political}	&	US	&	County borders	&	Advertising and vote shares	& Many\\
                \citet*{snider2015barriers}	&	US	&	Distance from airports	&	Airfares	& Many \\
				\citet*{Urquiola:2006}	&	Bolivia	&	Cohort size	&	Test scores	& 2\\
				\citet*{Urquiola:2009}	&	Chile	&	Cohort size	&	Test scores	& 3 \\
				\citet*{Van:2002}	&	U.S.	&	Aid score	&	Financial aid	& Many \\
				\citet*{Van:2008}	&	U.S.	&	Poverty Score	&	School aid	& Many \\
				\bottomrule
			\end{tabular}}
		\end{center}
	}
    {\scriptsize \textbf{Note}: ``Many'' refers to examples where either a large number of cutoff points are present or a continuum of cutoff points can be generated (e.g., the cutoff is a continuous random variable). This table excludes a large number of political science and related applications reported in \citet[Supplemental Appenedix]{Cattaneo-Keele-Titiunik-VazquezBare_2016_JOP}.}
\end{table}

\section{Additional Empirical Results}\label{sec:robustness}

In this section we explore the sensitivity of our empirical results when restricting the sample to the period 2007-2010. This check allows us to reduce the heterogeneity of the sample due to variation over time. The results are shown in Table \ref{tab:robustness}. We find overall similar results, with a positive and significant extrapolated effect of 12.8 percentage points. 

\begin{table}
	\begin{center}
		\caption{Results for ACCES loan eligibility on Post-Education Enrollment in restricted ample \label{tab:robustness}}
		\resizebox{.8\textwidth}{!}{\input{\pathin/table_results_650_short.txt}}
	\end{center}
	\footnotesize Notes. Local polynomial regression estimation with MSE-optimal bandwidth selectors and robust bias corrected inference. See \cite{Calonico-Cattaneo-Titiunik_2014_ECMA} and \cite{Calonico-Cattaneo-Farrell_2018_JASA} for methodological details, and \cite{Calonico-Cattaneo-Farrell-Titiunik_2017_Stata} and \cite{Cattaneo-Titiunik-VazquezBare_2020_Stata} for implementation. ``Eff. $N$'' indicates the effective sample size, that is, the sample size within the MSE-optimal bandwidth. ``Bw'' indicates the MSE-optimal bandwidth.
\end{table}

\begin{table}[h]
	\caption{Segments and Corresponding Parameters in Figure \ref{fig:estimands}}
	\label{tab:FigParams}
	\centering
	\resizebox{1\textwidth}{!}{
		\begin{tabular}{ccl}
			\toprule
			\textbf{Segment} & \textbf{Parameter} & \textbf{Description}\\  
			\midrule 
			$\overline{ab}$ & $\tau_\L(\x) = \underbrace{\mu_{1,\L}(\x)}_\text{Observable}  - \underbrace{\mu_{0,\L}(\x)}_\text{Unobservable}$
			& Extrapolation parameter of interest\vspace{0.2in} \\
			$\overline{bc}$ & $B(\x) = \underbrace{\mu_{0,\L}(\x)}_\text{Unobservable} - \underbrace{\mu_{0,\H}(\x)}_\text{Observable}$
			& Control facing $\L$ vs. control facing $\H$, at $X_i=\x$ \vspace{0.2in} \\
			$\overline{ac}$ & $\tau_\L(\x) + B(\x) = \underbrace{\mu_{1,\L}(\x)}_\text{Observable} - \underbrace{\mu_{0,\H}(\x)}_\text{Observable}$
			& Treated facing $\L$ vs. control facing $\H$, at $X_i=\x$ \vspace{0.2in} \\
			$\overline{ed}$ & $B(\L) = \underbrace{\mu_{0,\L}(\L)}_\text{Observable} - \underbrace{\mu_{0,\H}(\L)}_\text{Observable}$
			& Control facing $\L$ vs. control facing $\H$, at $X_i=\L$ \vspace{0.05in}\\
			\bottomrule
		\end{tabular}
	}
\end{table}

\section{Simulation Setup}

We provide further details on the simulation setup used in the paper. Potential outcome regression functions are generated in the following way:
\[\mu_{0,\H}(x) = \mathbf{r}_p(x)'\boldsymbol{\gamma},\quad \mu_{0,\ell}(x) = \mu_{0,\H}(x) + \Delta,\quad \mu_{1,c}(x) = \mu_{0,c}(x) + \tau \]
where $\mathbf{r}_p(x)$ is a $p$-th order polynomial basis for $x$. We set $p=4$. The value of $\boldsymbol{\gamma}$, given below, is chosen by running a regression of the observed outcome for the high-cutoff on a fourth order polynomial using data from our empirical application.
The observed variables are generated according to the following model:
\begin{align*}
X_i &\sim \mathsf{Uniform}(-1000,-1)\\
\{C_i\}_{i=1}^N &\sim \mathsf{FixedMargins}(N,N_{\ell}) \\
D_i &= \1(X_i\ge C_i)\\
Y_i &= \mu_{0,\H}(X_i) + \tau D_i + \Delta \1(C_i=\ell) + \mathsf{Normal}(0,\sigma^2)
\end{align*}
where $\mathsf{FixedMargins}(N,N_\ell)$ denotes a fixed-margins distribution that assigns $N_\ell$ observations out of the total $N$ sample to face cutoff $\ell$ and the remaining $N-N_{\ell}$ ones to face cutoff $\H$.

Proceeding as above, we set the parameters of the data generating process as follows:
\begin{align*}
\boldsymbol{\gamma}&=(-14.089,-0.074,-1.372e(-4),-1.125e(-07),-3.444e(-11))' \\
\Delta &= -0.14, \qquad \tau = 0.19, \qquad \sigma^2 = 0.3^2 \\
N &\in \{1000,2000,5000\},\quad N_\ell = N/2\\
\ell & = -850,\quad \H = -571, \quad \bar{x}=-650.
\end{align*}
All these parameters were estimated using local polynomial and related methods.

\section{Extensions and Generalizations}

We briefly discuss some extensions and generalizations of our main extrapolation results.

\subsection{Conditional-on-covariates Constant Bias}

Following \citet{Abadie_2005_RES}, we can relax the constant bias assumption to hold conditionally on observable characteristics. Let the bias term conditional on a vector of observable covariates $Z_i=z\in\mathcal{Z}$ be denoted by $B(x,z)=\E[Y_i(0)| X_i=x,Z_i=z,C_i=\L]-\E[Y_i(0)| X_i=x,Z_i=z,C_i=\H]$. Let $p(x,z)=\P(C_i=\ell| X_i=x,Z_i=z)$ denote the low-cutoff propensity score and $p(x)=\P(C_i=\ell|X_i=x)$.

We impose the following conditions:

\begin{assumption}[Ignorable Constant Bias]\label{assu:condcov}$ $
	\begin{enumerate}
		\item Conditional bias: $B(\L,z)=B(x,z)$ for all $x\in(\L,\H)$ and for all $z\in\mathcal{Z}$.
		\item Common support: $\delta<p(x,z)<1-\delta$ for some $\delta>0$, $x\in\mathcal{X}$ and for all $z\in\mathcal{Z}$.
		\item Continuity: $p(x,z)$ is continuous in $x$ for all $z\in\mathcal{Z}$.
	\end{enumerate}
\end{assumption}
Part 1 states that the selection bias is equal across cutoffs after conditioning on a covariates, part 2 is the usual assumption rulling out empty cells defined by the covariates $Z_i$, and part 3 assumes that the propensity score is continuous in the running variable. Then, letting $S_i=\1(C_i=\L)$, we have the following result, which is proven in the supplemental appendix.

\begin{theorem}[Covariate-Adjusted Extrapolation]\label{thm:condcov}
	Under Assumption 1 and \ref{assu:condcov},
	\begin{align*}
		\tau_\L(\bar{x})&=\E\left[\left.\frac{Y_iS_i}{p(\bar{x})}\right\vert X_i=\H\right]-\E\left[\left.\frac{Y_i(1-S_i)}{1-p(\bar{x},Z_i)}\cdot \frac{p(\bar{x},Z_i)}{p(\bar{x})}\right\vert X_i=x\right]\\
		&\quad +\E\left[\left.\frac{Y_i(1-S_i)p(\bar{x},Z_i)}{p(\bar{x})(1-p(\L,Z_i))}\cdot \frac{f_{X|Z}(\H|Z_i)}{f_{X|Z}(\L|Z_i)}\cdot \frac{f_X(\L)}{f_X(\H)}\right\vert X_i=\L\right] \\
		&\quad -\lim_{x\to \L^-}\E\left[\left.\frac{Y_iS_i p(\bar{x},Z_i)}{p(\bar{x})p(\L,Z_i)}\cdot\frac{f_{X|Z}(\H|Z_i)}{f_{X|Z}(\L|Z_i)}\cdot \frac{f_X(\L)}{f_X(\H)}\right\vert X_i=x\right].
	\end{align*}
\end{theorem}

This result is somewhat notationally convoluted, but it is straightforward to implement. It gives a precise formula for extrapolating RD treatment effects for values of the score above the cutoff $\L$, based on a conditional-on-$Z_i$ constant bias assumption. All the unknown quantities in Theorem \ref{thm:condcov} can be replaced by consistent estimators thereof, under appropriate regularity conditions.

\subsection{Non-Constant Bias}

Although the constant bias restriction (Assumption 2) is intuitive and allows for a helpful analogy with the difference-in-differences design, the RD setup leads for a natural extension via local polynomial extrapolation. The score is a continuous random variable and, under additional smoothness conditions of the bias function $B(x)$, we can replace the constant bias assumption with the assumption that $B(x)$ can be approximated by a polynomial expansion of $B(\L)$ around $x \in (\L,\H)$.

For example, using a polynomial of order one, we can approximate $B(x)$ at $\x$ as 
\begin{align}\label{eq:biasp1}
	B(\x) \approx B(\L) + \dot B(\L)\cdot[\x-\L]
\end{align}
where $\dot B(x)= \dot\mu_{0,\L}(x)-\dot\mu_{0,\H}(x)$ and $\dot\mu_{d,c}(x)=\partial \mu_{d,c}(x)/\partial x$. This shows that the constant bias assumption $B(x) = B(\L)$ can be seen as a special case of the above approximation, where the first derivatives of $\mu_{0,\L}(x)$ and $\mu_{0,\H}(x)$ are assumed equal to each other at $x=\x$. In contrast, the approximation in (\ref{eq:biasp1}) allows these derivatives to be different, and corrects the extrapolation at $\x$ using the difference between them at the point $\L$.

This idea allows, for instance, for different slopes of the control regression functions at $\x$, leading to $B(\x) \neq B(\L)$. The linear adjustment in expression (\ref{eq:biasp1}) provides a way to correct the bias term to account for the difference in slopes at the low cutoff $\L$. This represents a generalization of the constant assumption, which allows the intercepts of $\mu_{0,\L}(x)$
and $\mu_{0,\H}(x)$ to differ, but does not allow their difference to be a function of $x$. It is straightforward to extend this reasoning to employ higher order polynomials to approximate $B(\x)$, at the cost of a stronger smoothness and extrapolation assumptions.

\begin{assumption}[Polynomial-in-Score Bias]\label{assu:polybias} \hfill
	\begin{enumerate}
		\item \textit{Smoothness}: $\mu_{d,c}(x)=\E[Y_i(d)\mid X_i=x,C_i=c]$ are $p$-times continuously differentiable at $x=c$ for all $c\in\mathcal C $, $d=0,1$ and for some $p\in\{0,1,2,...\}$.
		\item \textit{Polynomial Approximation}: there exists a $p\in\{0,1,2,...\}$ such that, for $x \in (\L,\H)$
		\[B(x) =\sum_{s=0}^p \frac{1}{s!}B^{(s)}(\L)\cdot [x-\L]^s\]
		where $B^{(s)}(x)= \mu^{(s)}_{0,\L}(x)-\mu^{(s)}_{0,\H}(x)$ and $\mu^{(s)}_{d,c}(x)=\partial^s \mu_{d,c}(x)/\partial x^s$.
		
	\end{enumerate}
\end{assumption}
The main extrapolation result in can be generalized as follows.

\begin{theorem}\label{thm:extrapol2}
	Under Assumption \ref{assu:polybias}, for $\x \in (\L,\H )$,
	\[\tau_{\L}(\x) = \mu_{\L}(\x) -  \Big[\mu_{\H}(\x) + \sum_{s=0}^p \frac{1}{s!}B^{(s)}(\L)\cdot [x-\L]^s\Big].\]
\end{theorem}

This result establishes valid extrapolation of the RD treatment effect away from the low cutoff $\L$. This time the extrapolation is done via adjusting for not only the constant difference between the two control regression functions but also their higher-order derivatives. Heuristically, this result justifies approximating the control regression functions by a higher-order polynomial, local to the cutoff, and then using the additional information about higher-derivatives to extrapolate the treatment effects.

\subsection{Many Cutoffs}

All the identification results in the main paper hold for any number of cutoffs $J\ge 2$. The key issue to be assessed in this case is whether the constant bias assumption holds for all subpopulations. When this is the case, $\tau_c(x)$ is overidentified (for $x\ge c$), since there are many control groups that can be used to identify this parameter. In this setting, joint estimation can be performed using fixed effects models, as explained in the next section.

Furthermore, when more than two cutoffs are available, more control regression functions are therefore available for extrapolation. These could be combine to extrapolate or, alternatively, each of them could be used to extrapolate the RD treatment effect and only after these treatment effects could be combined. We relegate this interesting problem for future work.

\subsection{Multi-Scores and Geographic RD Designs}

\begin{figure}
	\centering
	\caption{Illustration of Multi-Score and Geographic RD Designs}
	\label{fig:RD-multiscore}
	\vspace{-.05in}
	\begin{subfigure}[b]{0.49\textwidth}
		\includegraphics[width=1\textwidth]{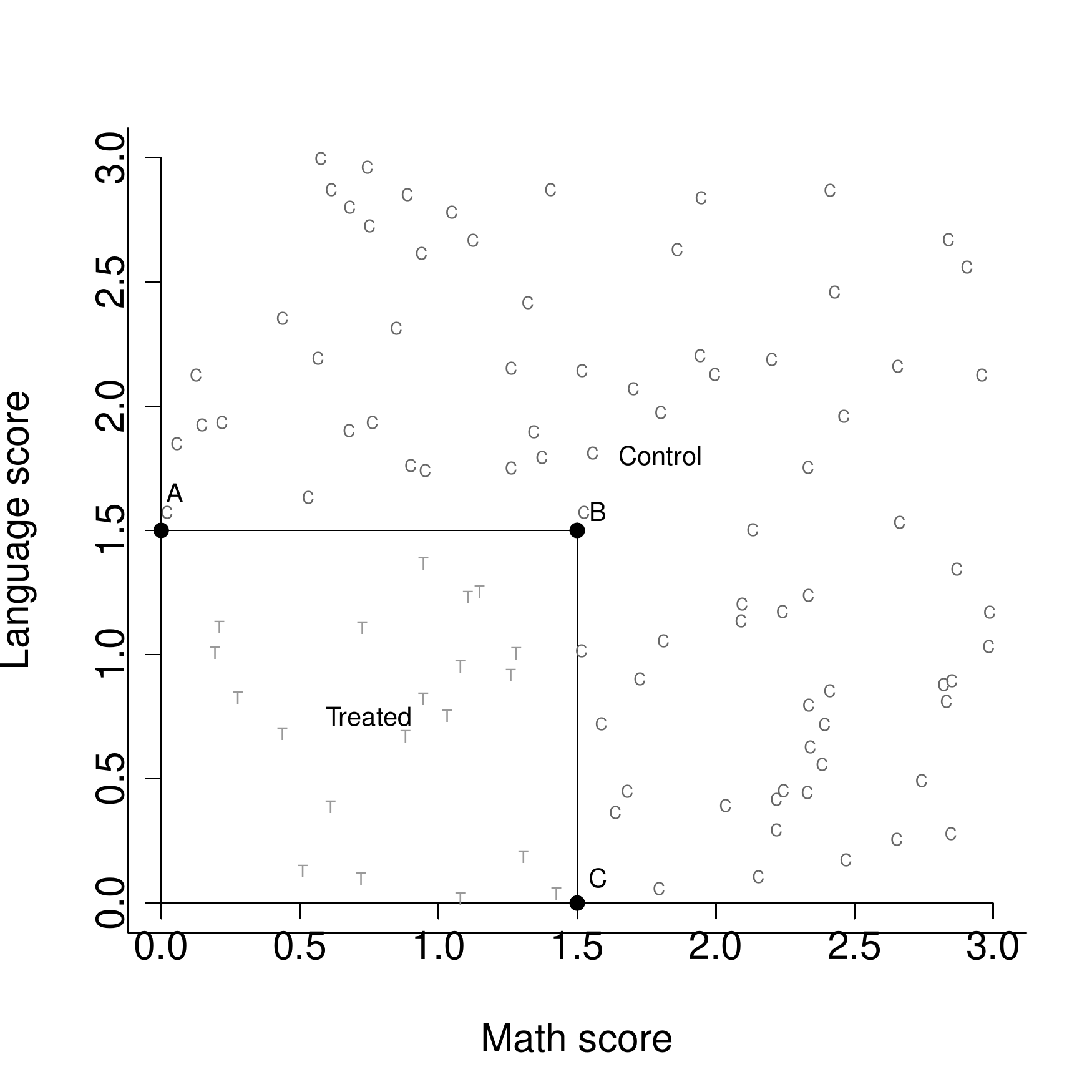}
		\caption{Multi-Score RD Design}
	\end{subfigure}	
	\hspace{-0.2in}
	\begin{subfigure}[b]{0.49\textwidth}
		\includegraphics[width=1\textwidth]{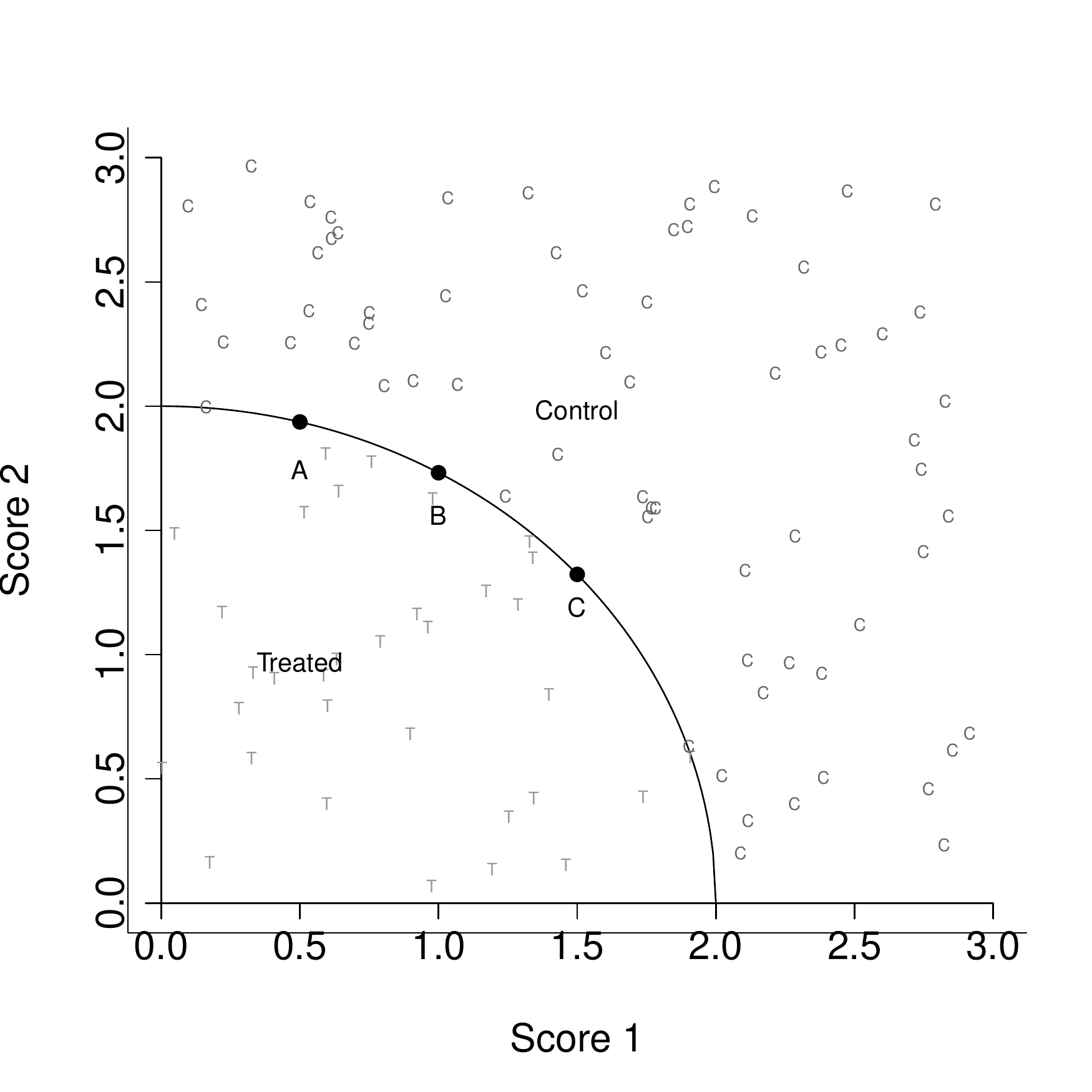}
		\caption{Geographic RD Design}
	\end{subfigure}	
\end{figure}

In many applications, RD designs include multiple scores \citep[e.g.,][]{PapayWillettMurnane2011-JoE,ReardonRobinson2012-JREE,Keele-Titiunik_2015_PA}. Examples include treatments assigned based on not exceeding a given threshold on two different scores (Figure \ref{fig:RD-multiscore}(a)) or based on being one side of some generic boundary (Figure \ref{fig:RD-multiscore}(b)). While these designs induce a continuum of cutoff points, it is usually better to analyze them using a finite number of cutoffs along the boundary determining treatment assignment.

For example, in Figure \ref{fig:RD-multiscore} we illustrate two settings with three chosen cutoff points (A, B, and C). In panel (a), the three cutoffs correspond to ``extremes'' over the boundary, while in panel (b) the cutoff points are chosen towards the ``center''. Once these cutoffs points are chosen, the analysis can proceed as discussed in the main paper. Usually, the multidimensionality of the problem is reduced by relying on some metric that maps the multi-score feature of the design to a unidimensional problem. For instance, $X_i$ is usually taken to be a measure of distance relative to the desired cutoff point. Once this mapping is constructed, all the ideas and methods discussed in the paper can be extended and applied to extrapolate RD treatment effects across cutoff points in multi-score RD designs.

\section{Relationship with Fixed Effects Models}

Consider a separable model for the potential outcomes and only two cutoffs $\L<\H$:
\begin{align*}
Y_i(0)&=g_0(X_i)+\gamma_0\mathbbm 1(C_i=\H)+\varepsilon_{0i},\quad \E[\varepsilon_{0i}\mid X_i,C_i]=0\\
Y_i(1)&=g_1(X_i)+\gamma_1\mathbbm 1(C_i=\H)+\varepsilon_{1i},\quad \E[\varepsilon_{1i}\mid X_i,C_i]=0
\end{align*}
This model assumes that $X_i,C_i$ and the error terms are separable. A key implication of separability is the absence of interaction between the cutoff and the score. In other words, changing the cutoff only shifts the conditional expectation function without affecting the slope. The above model implies:
\[\tau_c(x)=\E[Y_{i}(1)-Y_{i}(0)\mid X_i=x,C_i=c]=g_1(x)-g_0(x)+(\gamma_1-\gamma_0)\mathbbm 1(c=\H)\]

Also, let $D_i=\mathbbm 1(X_i\ge C_i)$, $S_i=\mathbbm 1(C_i=\H)$, $\gamma\equiv \gamma_0$ and $\delta\equiv \gamma_1-\gamma_0$, so that defining the observed outcome in the usual way, $Y_i=D_iY_{i}(1)+(1-D_i)Y_{i}(0)$, we get:
\[Y_i=g_0(X_i)+\gamma S_i+(g_1(X_i)-g_0(X_i))D_i+\delta D_i\times S_i+\varepsilon_i\]
where $\varepsilon_i=\varepsilon_{0i}D_i+\varepsilon_{1i}(1-D_i)$ and $\E[\varepsilon_i\mid X_i,C_i]=0$.

Although restrictive, this separable model nests several particular cases that are commonly used in RD estimation. For instance, if we assume:
\[g_0(x)=\alpha_0+\beta_0 x,\quad g_1(x)=\alpha_1+\beta_1 x,\]
the model reduces to:
\[Y_i=\alpha_0+\beta_0 X_i+(\alpha_1-\alpha_0)D_i +(\beta_1-\beta_0)X_i\times D_i+\gamma S_i+\delta D_i\times S_i+\varepsilon_i\]
which is the usual linear model with an interaction between the score and the treatment, with two additional terms that account for the presence of two cutoffs.

The assumption of separability is sufficient for the selection bias to be constant across cutoffs. More precisely,
\begin{align*}
\E[Y_i(0)\mid X_i=x,C_i=\L]&=g_0(x)\\
\E[Y_i(0)\mid X_i=x,C_i=\H]&=g_0(x)+\gamma_0\\
\Rightarrow B(x)&=\gamma_0,\quad \forall\, x
\end{align*}

In this setting, consider the case of many cutoffs is relatively straightforward. With $C_i\in\{c_0,c_1,...c_J\}:=\mathcal C$, the potential outcomes can be defined as:
\begin{align*}
Y_i(0)&=g_0(X_i)+\sum_{j=1}^J\gamma_{0j}\mathbbm 1(C_i=c_j)+\varepsilon_i\\
Y_i(1)&=g_{1j}(X_i)+\sum_{j=1}^J\gamma_{1j}\mathbbm 1(C_i=c_j)+\varepsilon_i
\end{align*}
with observed outcome:
\[Y_i=g_0(X_i)+\sum_{j=1}^J\gamma_{0j}S_{ij}+(g_{1j}(X_i)-g_0(X_i))D_i+\sum_{j=1}^J\delta_j D_i\times S_{ij}+\varepsilon_i\]
where $S_{ij}=\mathbbm 1(C_i=c_j)$ and $\delta_j=\gamma_{1j}-\gamma_{0j}$. As before, the above model implies that the selection bias does not depend on the running variable:
\[B(x,c_j,c_k)=\gamma_{0j}-\gamma_{0k}\]

The equation for the observed outcome can be rewritten as:

\[Y_{ij}=\gamma_j+g(X_{ij})+\theta_jD_{ij}+\tau_j(X_{ij})D_{ij}+\varepsilon_{ij}\]
where $Y_{ij}$ is the outcome for unit $i$ exposed to cutoff $c_j$ and $D_{ij}$ is equal to one if unit $i$ exposed to cutoff $c_j$ is treated, $D_{ij}=\mathbbm 1(X_i\ge c_j)\times \mathbbm 1(C_i=c_j)$. This model is similar to a one-way fixed effects model. 

\subsection{Example: linear case}

Using the above ``fixed-effects notation'', suppose $g_0$ and $g_1$ are linear:
\begin{align*}
Y_{ij}(0)&=\gamma_{0j}+\beta_0 X_{ij}+\varepsilon^0_{ij}\\
Y_{ij}(1)&=\gamma_{1j}+\beta_{1j} X_{ij}+\varepsilon^1_{ij}
\end{align*}
where $Y_{ij}(d)$ is the potential outcome of unit $i$ facing cutoff $c_j$ under treatment $d$. As before, the fact that the slopes change with the treatment status but not with the cutoff is implied by separability between the score and the cutoff indicator. The observed outcome is:

\[Y_{ij}=\gamma_{0j}+\beta_0X_{ij}+(\gamma_{1j}-\gamma_{0j})D_{ij}+(\beta_{1j}-\beta_0)X_{ij}\times D_{ij}+\varepsilon_{ij}\]
with $\varepsilon_{ij}=\varepsilon^1_{ij}D_{ij}+\varepsilon^0_{ij}(1-D_{ij})$.

Reparameterizing the model gives the estimating equation:

\[Y_{ij}=\gamma_j+\beta X_{ij}+\delta_jD_{ij}+\theta_j X_{ij}\times D_{ij}+\varepsilon_{ij}\]
which is a linear model including cutoff fixed effects, the running variable, the treatment variable with cutoff-varying coefficients and the interaction between the score and the treatment. In other words, this is the standard linear RD specification, but with different intercepts and slopes at each cutoff. Note that the coefficients for $X_{ij}$ do not vary with $j$. This captures the restriction that the slopes of the conditional expectations under no treatment are the same across cutoffs and also leads to a straightforward specification test.

Under the linear specification, the treatment effect evaluated at $c_k$ for units facing cutoff $c_j$ with $c_k>c_j$ is given by:
\[\tau_{c_j}(\bar{x})=\gamma_{1j}-\gamma_{0j}+(\beta_{1j}-\beta_0)\bar{x}\]
and can be estimated as:
\[\hat\tau_{c_j}(\bar{x})=\hat\delta_j+\hat\theta_j \bar{x}\]
or simply as $\hat\delta_j$ if the running variable is appropriately re-centered.

Clearly, assuming a linear specification would in principle allow one to estimate the effects not only at the cutoffs but also at any other point in the range of the score. The treatment effects away from the cutoffs, however, are not identified nonparametrically and their identification relies purely on the functional form assumption, which can make them less credible.

The specification test mentioned above for the linear case consists on running the regression:
\[Y_{ij}=\gamma_j+\beta_j X_{ij}+\delta_jD_{ij}+\theta_j X_{ij}\times D_{ij}+\varepsilon_{ij}\]
and testing:
\[H_0:\quad \beta_1=\beta_2=...=\beta_J.\]


\section{Local Randomization Methods}

Our core extrapolation ideas can be adapted to the local randomization RD framework of \cite{Cattaneo-Frandsen-Titiunik_2015_JCI} and \cite{Cattaneo-Titiunik-VazquezBare_2017_JPAM}. In this alternative framework, only units whose scores lay within a fixed (and small) neighborhood around the cutoff  are considered, and their potential outcomes are regarded as fixed. The key source of randomness comes from the treatment assignment mechanism---the probability law placing units in control and treatment groups, and consequently the analysis proceeds as if the RD design was a randomized experiment within the neighborhood. See \cite{Rosenbaum_2010_Book} and \cite{Imbens-Rubin_2015_Book} for background on classical Fisher and Neyman approaches to the analysis of experiments.

As in the continuity-based approach, the Multi-Cutoff RD design can be analyzed using local randomization methods by either normalizing-and-pooling all cutoffs or by studying each cutoff separately. However, to extrapolate away from an RD cutoff (i.e., outside the small neighborhood where local randomization is assumed to hold), further strong identifying assumptions are needed. To discuss these additional assumptions, we first formalize the local randomization (LR) framework for extrapolation in a Multi-Cutoff RD design.

Recall that $\mathcal{C}=\{\L,\H\}$ for simplicity. Let $\mathcal{N}_x$ be a (non-empty) LR neighborhood around $x\in[\L,\H]$, that is,
\[\mathcal{N}_x = [x-w , x+w],\qquad w>0,\]
where the (half) window length may be different for each center point $x$. The other neighborhoods discussed below are defined analogously; for example, $\mathcal{N}_{\x} = [\x-w , \x+w]$ for some $w>0$, possibly different across neighborhoods. Let also $y_{ic}(d,x)$ be a non-random potential outcome for unit $i$ when facing cutoff $c$ with treatment assignment $d$ and running variable $x$. Consequently, in this Multi-Cutoff RD LR framework each unit has non-random potential outcomes $\{y_{i\L}(0,x),y_{i\L}(1,x),y_{i\H}(0,x),y_{i\H}(1,x)\}$, for each $x\in\mathcal{X}$. The observed outcome is $Y_i = y_{iC_i}(0,X_i) \mathbbm{1}(X_i< C_i) + y_{iC_i}(1,X_i) \mathbbm{1}(X_i\ge C_i)$ where $C_i\in\mathcal{C}$. Recall that our goal is to extrapolate the RD treatment effect to a point $\x\in(\L,\H]$ on the support of the running variable.

\begin{assumption}[LR Extrapolation]\label{assu:LR} $ $
	
	\begin{itemize}
		\item[\emph{(i)}] For all $i$ such that $X_i\in\mathcal{N}_\L$,  \[\{y_{i\L}(0,x),y_{i\L}(1,x),y_{i\H}(0,x)\}=\{y_{i\L}(0,\L),y_{i\L}(1,\L),y_{i\H}(0,\L)\}\] for all $x\in\mathcal{N}_\L$, and are non-random. Furthermore, the treatment assignment mechanism is known.
		
		\item[\emph{(ii)}] For $\x\in(\L,\H]$ such that $\mathcal{N}_\L\cap\mathcal{N}_{\x}=\emptyset$ and for all $i$ such that $X_i\in\mathcal{N}_{\x}$,  \[\{y_{i\L}(0,x),y_{i\L}(1,x),y_{i\H}(0,x)\}=\{y_{i\L}(0,\x),y_{i\L}(1,\x),y_{i\H}(0,\x)\}\] for all $x\in\mathcal{N}_{\x}$, and are non-random. Furthermore, the treatment assignment mechanism is known.
		
		\item[\emph{(iii)}] There exists a constant $\Delta\in\mathbb{R}$ such that: \[y_{i\L}(0,\L) = y_{i\H}(0,\L) + \Delta,\qquad \text{for all $i$ such that} X_i\in\mathcal{N}_\L,\] and
	    \[y_{i\L}(0,\x) = y_{i\H}(0,\x) + \Delta, \qquad \text{for all $i$ such that } X_i\in\mathcal{N}_{\x}.\]
		
	\end{itemize}
\end{assumption}

Assumption \ref{assu:LR}(i) is analogous to Assumption 1 in \cite{Cattaneo-Frandsen-Titiunik_2015_JCI} applied to the RD cutoff $c=\L$, except for the presence of one additional potential outcome, $y_{i\H}(0)$, which we will use for extrapolation in the Multi-Cutoff RD design. Likewise, Assumption \ref{assu:LR}(ii) postulates the existence of a LR neighborhood for the desired point of extrapolation $\x$. Finally, Assumption \ref{assu:LR}(iii) imposes a relationship between (the difference of) control potential outcomes in the LR neighborhood of $c=\L$ and (the difference of) control potential outcomes in the LR neighborhood of $c=\x$, which we will use to impute the missing control potential outcomes for units exposed to the low cutoff $c=\L$ but with scores within $\mathcal{N}_{\x}$. To conserve space and notation, we do not extend Assumption \ref{assu:LR} to allow for regression adjustments within the LR neighborhoods as in \cite{Cattaneo-Titiunik-VazquezBare_2017_JPAM}, but we do include the corresponding results in our empirical application.

In the LR framework, extrapolation requires imputing both the assignment mechanism and the missing control potential outcomes $y_{i\L}(0,\bar{x})$ within $\mathcal{N}_{\x}$. As a consequence, extrapolation beyond the standard LR neighborhood $\mathcal{N}_{\L}$ requires very strong assumptions. Assumption \ref{assu:LR} provides a set of conditions that lead to valid extrapolation. The parameter of interest is the average effect of the treatment for units with $X_i\in\mathcal{N}_{\bar{x}}$: 
\[\tau_{\mathsf{LR}} = \frac{1}{N_{\bar{x}}}\sum_{X_i\in\mathcal{N}_{\bar{x}}} (y_{i\ell}(1,\bar{x})-y_{i\ell}(0,\bar{x}))\]
where $N_{\bar{x}}$ is the number of units inside the window $\mathcal{N}_{\x}$ around $\bar{x}$. Under Assumption \ref{assu:LR}, this parameter equals: $\frac{1}{N_{\bar{x}}}\sum_{X_i\in\mathcal{N}_{\bar{x}}}(y_{i\ell}(1,\bar{x})-y_{i\H}(0,\bar{x}))-\Delta$, which is identifiable from the data. We implement this result as follows. First, we construct an estimate of $\Delta$ as the difference-in-means for control units facing cutoffs $\ell$ and $\H$ with $X_i\in\mathcal{N}_\L$, which we denote by $\hat{\Delta}$:
\[\hat{\Delta}=\bar{Y}_{\ell}(0,\ell)-\bar{Y}_\H(0,\ell)\]
where
\[\bar{Y}_{\ell}(0,\ell)=\frac{1}{N^0_{\ell}(\ell)}\sum_{X_i\in \mathcal{N}_\ell} Y_i\1(C_i=\ell)(1-D_i),\quad \bar{Y}_\H(0,\ell)=\frac{1}{N_{\H}(\ell)}\sum_{X_i\in \mathcal{N}_\ell} Y_i\1(C_i=\H)\]
and 
\[N^0_{\ell}(\ell)=\sum_{X_i\in \mathcal{N}_\ell} \1(C_i=\ell)(1-D_i),\quad N_{\H}(\ell)=\sum_{X_i\in \mathcal{N}_\ell} \1(C_i=\H).\]
Second, we estimate the treatment effects as:
\[\hat{\tau}_{\mathsf{LR}}=\bar{Y}_\ell(1,\bar{x})-\bar{Y}_\H(0,\bar{x})-\hat{\Delta}\]
where 
\[\bar{Y}_\ell(1,\bar{x})=\frac{1}{N_{\ell}(\bar{x})}\sum_{X_i\in \mathcal{N}_{\bar{x}}} Y_i\1(C_i=\ell),\quad \bar{Y}_\H(0,\bar{x})=\frac{1}{N_{\H}(\bar{x})}\sum_{X_i\in \mathcal{N}_{\bar{x}}} Y_i\1(C_i=\H)\]
and
\[N_{\ell}(\bar{x})=\sum_{X_i\in \mathcal{N}_{\bar{x}}} \1(C_i=\ell),\quad N_{\H}(\bar{x})=\sum_{X_i\in \mathcal{N}_{\bar{x}}} \1(C_i=\H).\]
Finally, for the assignment mechanism, we assume:
\[\P[C_i=\H|X_i\in\mathcal{N}_c]=\frac{N_\H(c)}{N_{\ell}(c)+N_\H(c)},\quad \forall\, i:X_i\in\mathcal{N}_c,\, c\in\{\ell,\bar{x}\}\]
and
\[\P[D_i=0|C_i=\ell,X_i\in \mathcal{N}_\ell]=\frac{N^0_{\ell}(\ell)}{N_{\ell}(\ell)},\quad N_{\ell}(\ell)=\sum_{X_i\in\mathcal{N}_\ell}\1(C_i=\ell)\]

It is straightforward to show that under this assignment mechanism and Assumption \ref{assu:LR}, $\hat{\Delta}$ and $\hat{\tau}$ are unbiased for their corresponding parameters. Our approach is not the only way to develop LR methods for extrapolation, but for simplicity we focus on the above construction which mimics closely the continuity-based proposed in the previous sections.

In this setting, inference can be conducted using Fisherian randomization inference by permuting the cutoff indicator $\1(C_i=\ell)$ on the adjusted outcomes $Y_i^A=Y_i+\Delta\1(C_i=\H)$ among units in $\mathcal{N}_{\bar{x}}$. However, the inference procedure needs to account for the fact that $\Delta$ is unknown and needs to be estimated. We propose two alternatives to deal with this issue. The first one, suggested by \citet{Berger-Boos_1994_JASA}, consists on constructing a $(1-\eta)$-level confidence interval for $\Delta$, $S_\eta$, and defining the p-value:
\[p^*(\eta) = \sup_{\Delta\in S_\eta}p(\Delta)+\eta\]
which can be shown to be valid in the sense that $\P[p^*(\eta) \le \alpha]\le \alpha$. 

Our second inference procedure, based on Neyman's sampling approach, consists on using the standard normal distribution to approximate the distribution of the studentized statistic:
\[T=\frac{\hat{\tau}}{\sqrt{V_1+V_{\hat{\Delta}}}}\]
where $V_1$ is the estimated variance of the difference in means $\bar{Y}_\ell(1,\bar{x})-\bar{Y}_\H(0,\bar{x})$ and $V_{\hat{\Delta}}$ is the estimated variance of $\hat{\Delta}$.

\subsection{Empirical Application}

We use our proposed LR methods to investigate the external validity of the ACCES program RD effects. As mentioned in the paper, our sample has observations exposed to two cutoffs, $\L = -850$ and $\H = -571$. We begin by extrapolating the effect to the point $\x = -650$; our focus is thus the effect of eligibility for ACCES on whether the student enrolls in a higher education program for the subpopulation exposed to cutoff 850 when their SABER 11 score is $650$.

Table \ref{tab:results_LR} presents empirical results using the local randomization framework. We construct the neighborhoods $\mathcal{N}_\ell$ and $\mathcal{N}_{\bar{x}}$ using the 50 closest observations to the evaluation point of interest. To calculate $p^*(\eta)$, we construct a $99$ percent confidence interval for $\Delta$ based on the normal approximation, which can be justified using large sample approximations in either a fixed potential outcomes model (Neyman) or a standard repeated sampling model (superpopulation). We estimate $\hat{\tau}_{\mathsf{LR}}$ using a constant model and using a linear adjustment \citep[see][for details]{Cattaneo-Titiunik-VazquezBare_2017_JPAM}. Overall, the results are very similar to the ones obtained using the continuity-based approach. We find positive effects of around 20 percentage points that are significant at the 5 percent level using either Fisherian-based or Neyman-based inference. 

\begin{figure}
	\centering
	\caption{Sensitivity Analysis for Local Randomization}\label{fig:LR_sensit}
	\begin{subfigure}[b]{0.45\textwidth}
		\includegraphics[width=\textwidth]{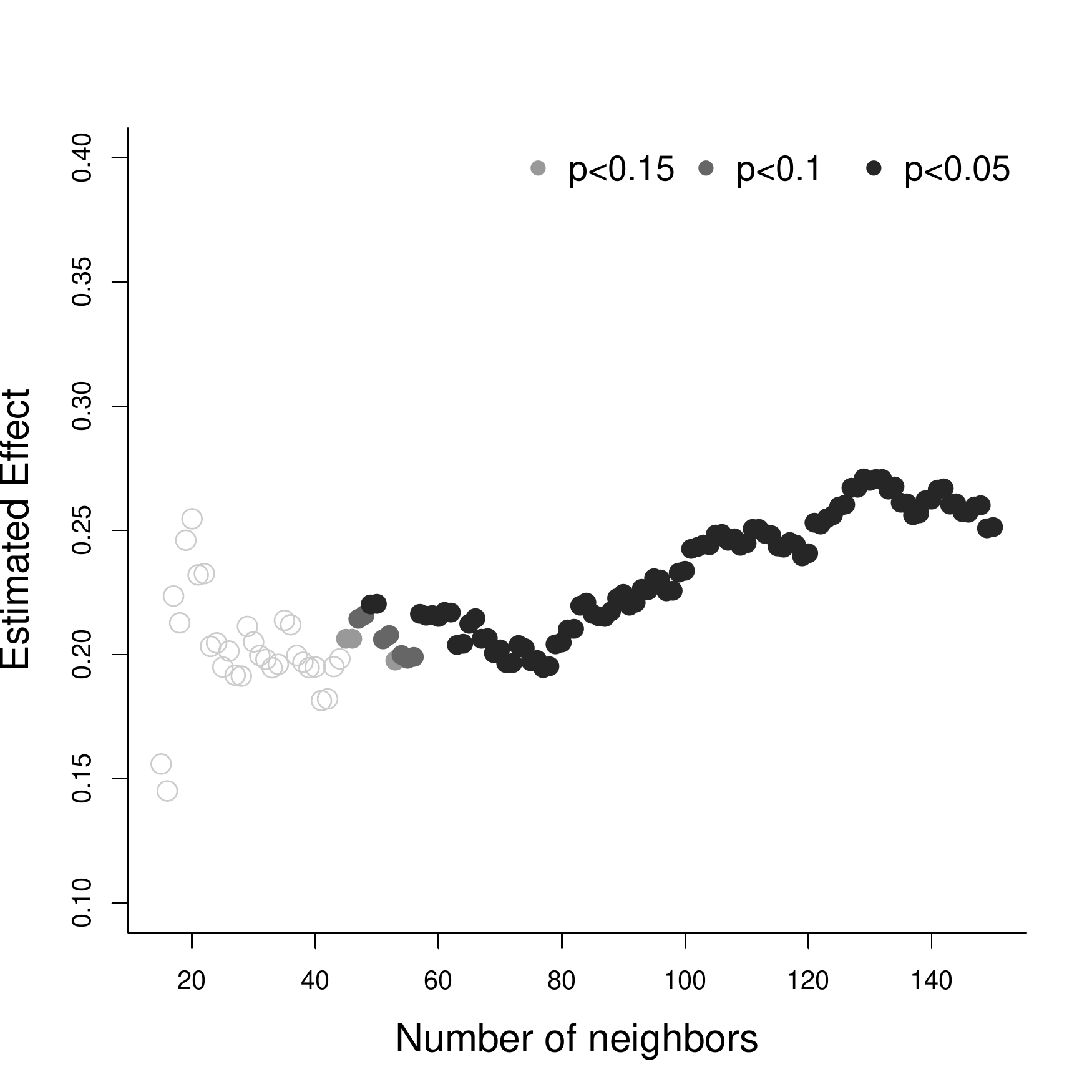}
		\caption{Constant Model \label{fig:sensit_cons}}
		\vspace{-.25in}
	\end{subfigure}
	\begin{subfigure}[b]{0.45\textwidth}
		\includegraphics[width=\textwidth]{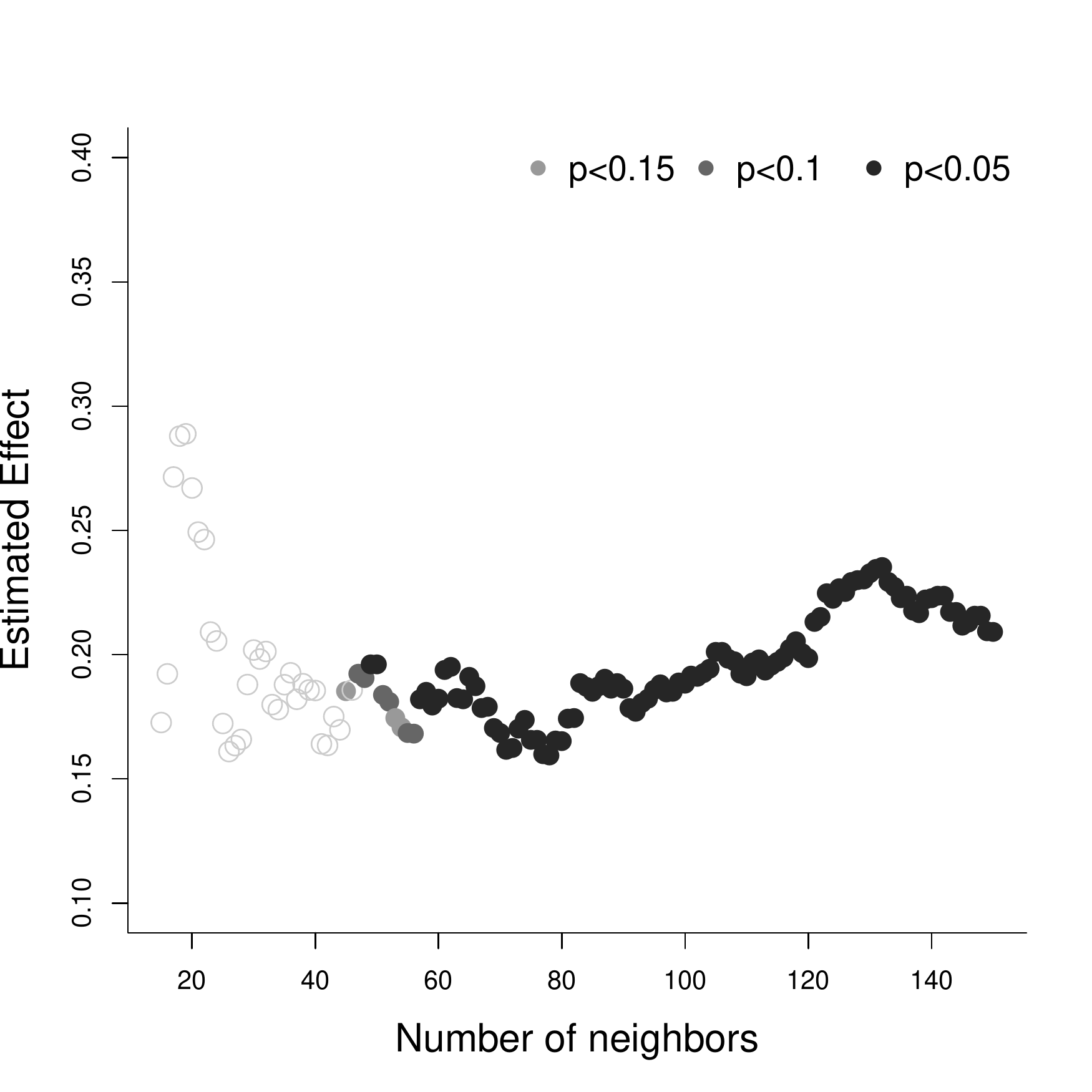}
		\caption{Linear Adjustment \label{fig:sensit_lin}}
		\vspace{-.25in}
	\end{subfigure}
	
	\begin{flushleft}\footnotesize Notes. The figure plots the estimated effect as a function of the number of nearest neighbors used around the cutoff for estimation. The left panel plots the estimates under a constant model. The right panel plots the estimates using a linear adjustment model. Hollow markers indicate p-value $\ge 0.15$. Light gray markers indicate p-value $<0.15$. Dark gray markers indicate p-value $<0.1$. Black markers indicate p-value $<0.05$. Randomization inference p-values constructed using the \cite{Berger-Boos_1994_JASA} method. \end{flushleft}
\end{figure}

To assess robustness of the LR methods, Figure \ref{fig:LR_sensit} shows how the estimated effect and its corresponding randomization inference p-value change when varying the number of nearest neighbors used to construct $\mathcal{N}_\ell$ and $\mathcal{N}_{\bar{x}}$. The magnitude of the estimated effect remains stable when increasing the length of the window, particularly for the linear adjustment case which can help to reduce bias when the corresponding regression functions are not constant. In terms of inference, while the p-values we construct can be very conservative, we find significant effects at the 5 percent level when using around 45 observations in each neighborhood.

\clearpage

\begin{table}
	\begin{center}
		\caption{Empirical Results under Local Randomization \label{tab:results_LR}}
		\resizebox{\textwidth}{!}{\input{\pathin/table_results_LR.txt}}
	\end{center}
	\footnotesize Notes. Estimated effect under the local randomization framework. Randomization inference p-values for $\hat{\tau}_{\mathsf{LR}}$ constructed using the \cite{Berger-Boos_1994_JASA} method. Neyman-based p-values constructed using a large-sample normal approximation. Estimates calculated using a constant model based on difference in meansand a linear regression adjustment \citep[see][for details]{Cattaneo-Titiunik-VazquezBare_2017_JPAM}. ``Eff. $N$'' indicates the effective sample size, that is, the sample size within the local randomization neighborhood. 
\end{table}

\clearpage


\section{Proofs and Derivations}

\subsection{Proof of Theorem 1}

Follows immediately under the stated assumptions by the derivation provided in the paper.

\subsection{Proof of Theorem 2}

 We omit the $i$ subscript to simplify notation. The average observed outcomes are:
\begin{align*}
\E[Y|X=x,C=c]&=\E[Y(1)D(x,c)+Y(0)(1-D(x,c))|X=x,C=c]\\
&=\E[(Y(1)-Y(0))D(x,c)|X=x,C=c]+\E[Y(0)|X=x,C=c]
\end{align*}
The difference in outcomes for units facing cutoffs $l<h$ at $X=x\in(l,h)$ is:
\begin{align*}
\Delta(x)&=\E[Y|X=x,C=l]-\E[Y|X=x,C=h]\\
&=\E[(Y(1)-Y(0))D(x,l)|X=x,C=l]-\E[(Y(1)-Y(0))D(x,h)|X=x,C=h] \\
&\quad +\E[Y(0)|X=x,C=l]-\E[Y(0)|X=x,C=h]
\end{align*}
Assuming parallel regression functions under no treatment, the double difference recovers:
\begin{align*}
&\Delta(x)-\Delta(l)\\
&=\E[Y|X=x,C=l]-\E[Y|X=x,C=h]-\lim_{x\uparrow l}\E[Y|X=x,C=l]+\E[Y|X=l,C=h]\\
&=\E[(Y(1)-Y(0))D(x,l)|X=x,C=l]-\E[(Y(1)-Y(0))D^-(l,l)|X=l,C=l] \\
&\quad -\{\E[(Y(1)-Y(0))D(x,h)|X=x,C=h]-\E[(Y(1)-Y(0))D(l,h)|X=l,C=h]\}
\end{align*}
where $D^-(l,l)=\lim_{x\uparrow l}D(x,l)$. Note that in a sharp design, $D(x,l)=1$ and $D(x,h)=D(l,h)=D^-(l,l)=0$ giving our previous result. After adding and subtracting,
\begin{align*}
&\Delta(x)-\Delta(l)\\
&=\E[Y|X=x,C=l]-\E[Y|X=x,C=h]-\lim_{x\uparrow l}\E[Y|X=x,C=l]+\E[Y|X=l,C=h]\\
&=\E[(Y(1)-Y(0))(D(x,l)-D^-(l,l))|X=x,C=l] \\
&\quad +\E[(Y(1)-Y(0))D^-(l,l)|X=x,C=l] -\E[(Y(1)-Y(0))D^-(l,l)|X=l,C=l] \\
&\quad -\{\E[(Y(1)-Y(0))D(x,h)|X=x,C=h]-\E[(Y(1)-Y(0))D(l,h)|X=l,C=h]\}
\end{align*}
Under one-sided non-compliance, units below the cutoff never get the treatment, so $D(x,h)=D(l,h)=D^-(l,l)=0$. In this case, 
\begin{align*}
\Delta(x)-\Delta(l)&=\E[(Y(1)-Y(0))D(x,l)|X=x,C=l]\\
&=\E[Y(1)-Y(0)|X=x,C=l,D(x,l)=1]\P[D(x,l)=1|X=x,C=l]
\end{align*}
It follows that
\begin{align*}
\frac{\Delta(x)-\Delta(l)}{\E[D|X=x,C=l]}&=\E[Y(1)-Y(0)|X=x,C=l,D(x,l)=1]
\end{align*}
which in this case equals the (local) average effect on the compliers.

\subsection{Proof of Theorem \ref{thm:condcov}}

Following analogous steps to the derivation for the unconditional case, we obtain that:
\begin{align*}
\E[\tau_i| X_i=\bar{x},Z_i=z,C_i=\L]&=\E[Y_i| X_i=\bar{x},Z_i=z,C_i=\L]\\
\quad & - \E[Y_i| X_i=\bar{x},Z_i=z,C_i=\H]-B(\L,z)
\end{align*}
and
\[\tau_\L(\bar{x})=\E\left\{\E[\tau_i| X_i=\bar{x},Z_i,C_i=\L]| X_i=\bar{x},C_i=\L\right\}\]

Define $p(x,z)=\P(C_i=\L| X_i=x,Z_i=z)$ and $S_i=\mathbbm{1}(C_i=\L)$. We have that:
\begin{align*}
\E[Y_i| X_i=\bar{x},C_i=\L,Z_i]&=\E\left[\left.\frac{Y_iS_i}{p(\bar{x},Z_i)}\right\vert X_i=\bar{x},Z_i\right]\\
\E[Y_i| X_i=\L,C_i=\H,Z_i]&=\E\left[\left.\frac{Y_i(1-S_i)}{1-p(\L,Z_i)}\right\vert X_i=\L,Z_i\right]
\end{align*}
and similarly for the remaining two terms, we obtain, assuming that $p(x,z)$ is continuous in $x$ (and limits can be interchanged),
\begin{align*}
\E[\tau_i | X_i=\bar{x},Z_i=z,C_i=\L]&=\E\left[\left.\frac{Y_iS_i}{p(\bar{x},Z_i)}\right\vert X_i=\bar{x},Z_i\right]-\E\left[\left.\frac{Y_i(1-S_i)}{1-p(\bar{x},Z_i)}\right\vert X_i=\bar{x},Z_i\right]\\
& +\E\left[\left.\frac{Y_i(1-S_i)}{1-p(\L,Z_i)}\right\vert X_i=\L,Z_i\right]-\lim_{x\to \L^-}\E\left[\left.\frac{Y_iS_i}{p(\L,Z_i)}\right\vert X_i=x,Z_i\right]
\end{align*}

To simplify the notation, let $\tau_c(x,z)=\E[\tau_i| X_i=x,C_i=c,Z_i=z]$, $\tau_c(x)=\E[\tau_i| X_i=x,C_i=c]$ and $p(x)=\P(C_i=\L| X_i=x)$. By Bayes' rule:
\begin{align*}
\tau_\L(\bar{x})&=\int \tau_\L(\bar{x},z) dF_{Z|X,C}(z|\bar{x},\L)\\
&=\int \tau_\L(\bar{x},z)\frac{p(\bar{x},z)}{p(\bar{x})}dF_{Z|X}(z|\bar{x})\\
&=\E\left[\left.\tau_\L(\bar{x},Z_i)\frac{p(\bar{x},Z_i)}{p(\bar{x})}\right\vert X_i=\bar{x}\right]
\end{align*}
Now replace $\tau_\L(\bar{x},Z_i)$ with its observed counterpart and split the outer expectation into the four summands. We have that:
\begin{align*}
\E\left[\E\left[\left.\left.\frac{Y_iS_i}{p(\bar{x},Z_i)}\right\vert X_i=\bar{x},Z_i\right]\frac{p(\bar{x},Z_i)}{p(\bar{x})}\right\vert X_i=\bar{x}\right]&=\E\left[\E\left[\left.\left.\frac{Y_iS_i}{p(\bar{x})}\right\vert X_i=\bar{x},Z_i\right]\right\vert X_i=\bar{x}\right]\\
&=\E\left[\left.\frac{Y_iS_i}{p(\bar{x})}\right\vert X_i=\bar{x}\right]
\end{align*}
and
\begin{align*}
&\E\left[\E\left[\left.\left.\frac{Y_i(1-S_i)}{1-p(\L,Z_i)}\right\vert X_i=\L,Z_i\right]\frac{p(\bar{x},Z_i)}{p(\bar{x})}\right\vert X_i=\bar{x}\right]\\
&\qquad=\E\left[\E\left[\left.\left.\frac{Y_i(1-S_i)}{1-p(\L,Z_i)}\cdot \frac{p(\bar{x},Z_i)}{p(\bar{x})}\right\vert X_i=\L,Z_i\right]\right\vert X_i=\bar{x}\right]\\
&\qquad=\E\left[\left.\frac{Y_i(1-S_i)}{p(\bar{x})}\cdot \frac{p(\bar{x},Z_i)}{1-p(\L,Z_i)}\cdot \frac{f_{Z|X}(Z_i|\bar{x})}{f_{Z|X}(Z_i|\L)}\right\vert X_i=\L\right] \\
&\qquad=\E\left[\left.\frac{Y_i(1-S_i)p(\bar{x},Z_i)}{p(\bar{x})(1-p(\L,Z_i))}\cdot \frac{f_{X|Z}(\bar{x}|Z_i)}{f_{X|Z}(\L|Z_i)}\cdot \frac{f_X(\L)}{f_X(\bar{x})}\right\vert X_i=\L\right] 
\end{align*}
Similarly, 
\begin{align*}
\E\left[\E\left[\left.\left.\frac{Y_i(1-S_i)}{1-p(\bar{x},Z_i)}\right\vert X_i=\bar{x},Z_i\right]\frac{p(\bar{x},Z_i)}{p(\bar{x})}\right\vert X_i=\bar{x}\right]&=\E\left[\left.\frac{Y_i(1-S_i)}{1-p(\bar{x},Z_i)}\cdot \frac{p(\bar{x},Z_i)}{p(\bar{x})}\right\vert X_i=x\right]
\end{align*}
and finally, under regularity conditions to interchange limits and expectations,
\begin{align*}
&\E\left[\lim_{x\to \L^-}\E\left[\left.\left.\frac{Y_iS_i}{p(\L,Z_i)}\right\vert X_i=x,Z_i\right]\frac{p(\bar{x},Z_i)}{p(\bar{x})}\right\vert X_i=\bar{x}\right]\\
&\qquad=\lim_{x\to \L^-}\E\left[\E\left[\left.\left.\frac{Y_iS_i\cdot p(\bar{x},Z_i)}{p(\bar{x})p(\L,Z_i)}\right\vert X_i=x,Z_i\right]\right\vert X_i=\bar{x}\right]\\
&\qquad=\lim_{x\to \L^-}\E\left[\left.\frac{Y_iS_i p(\bar{x},Z_i)}{p(\bar{x})p(\L,Z_i)}\cdot\frac{f_{Z|X}(Z_i|\H)}{f_{Z|X}(Z_i|\L)}\right\vert X_i=x\right]\\
&\qquad=\lim_{x\to \L^-}\E\left[\left.\frac{Y_iS_i p(\bar{x},Z_i)}{p(\bar{x})p(\L,Z_i)}\cdot\frac{f_{X|Z}(\H|Z_i)}{f_{X|Z}(\L|Z_i)}\cdot \frac{f_X(\L)}{f_X(\H)}\right\vert X_i=x\right]
\end{align*}

Putting all the results together,

\begin{align*}
\tau_\L(\bar{x})&=\E\left[\left.\frac{Y_iS_i}{p(\bar{x})}\right\vert X_i=\H\right]-\E\left[\left.\frac{Y_i(1-S_i)}{1-p(\bar{x},Z_i)}\cdot \frac{p(\bar{x},Z_i)}{p(\bar{x})}\right\vert X_i=x\right]\\
&\quad +\E\left[\left.\frac{Y_i(1-S_i)p(\bar{x},Z_i)}{p(\bar{x})(1-p(\L,Z_i))}\cdot \frac{f_{X|Z}(\H|Z_i)}{f_{X|Z}(\L|Z_i)}\cdot \frac{f_X(\L)}{f_X(\H)}\right\vert X_i=\L\right] \\
&\quad -\lim_{x\to \L^-}\E\left[\left.\frac{Y_iS_i p(\bar{x},Z_i)}{p(\bar{x})p(\L,Z_i)}\cdot\frac{f_{X|Z}(\H|Z_i)}{f_{X|Z}(\L|Z_i)}\cdot \frac{f_X(\L)}{f_X(\H)}\right\vert X_i=x\right].
\end{align*}

\subsection{Proof of Theorem \ref{thm:extrapol2}}

Follows immediately under the stated assumptions by the derivation provided in the paper.

\newpage\onehalfspacing

\bibliographystyle{jasa}
\bibliography{Cattaneo-Keele-Titiunik-VazquezBare_2020_JASA--bib}

%% file: inputs/table_results_650.txt
\begin{tabular}{lrcrcrcrc}
\hline\hline
\multicolumn{1}{l}{\bfseries }&\multicolumn{1}{c}{\bfseries }&\multicolumn{1}{c}{\bfseries }&\multicolumn{1}{c}{\bfseries }&\multicolumn{1}{c}{\bfseries }&\multicolumn{1}{c}{\bfseries }&\multicolumn{1}{c}{\bfseries }&\multicolumn{2}{c}{\bfseries Robust BC Inference}\tabularnewline
\cline{8-9}
\multicolumn{1}{l}{}&\multicolumn{1}{c}{Estimate}&\multicolumn{1}{c}{}&\multicolumn{1}{c}{Bw}&\multicolumn{1}{c}{}&\multicolumn{1}{c}{Eff. $N$}&\multicolumn{1}{c}{}&\multicolumn{1}{c}{p-value}&\multicolumn{1}{c}{95\% CI}\tabularnewline
\hline
{\bfseries RD effects}&&&&&&&&\tabularnewline
~~$C = -850$&0.137&&71.7&&71&&0.007&$[\;$0.036$\;,\;$0.232$\;]$\tabularnewline
~~$C = -571$&0.169&&136.3&&133&&0.103&$[\;$-0.039$\;,\;$0.428$\;]$\tabularnewline
~~Weighted&0.156&&&&204&&0.021&$[\;$0.025$\;,\;$0.314$\;]$\tabularnewline
~~Pooled&0.125&&147.6&&291&&0.029&$[\;$0.012$\;,\;$0.219$\;]$\tabularnewline
\hline
{\bfseries Naive difference}&&&&&&&&\tabularnewline
~~$\mu_{\ell}(-650)$&0.756&&240.1&&441&&&\tabularnewline
~~$\mu_h(-650)$&0.706&&131.2&&202&&&\tabularnewline
~~Difference&0.050&&&&&&0.179&$[\;$-0.020$\;,\;$0.107$\;]$\tabularnewline
\hline
{\bfseries Bias}&&&&&&&&\tabularnewline
~~$\mu_{\ell}(-850)$&0.525&&54.9&&54&&&\tabularnewline
~~$\mu_h(-850)$&0.667&&144.2&&230&&&\tabularnewline
~~Difference&-0.142&&&&&&0.004&$[\;$-0.274$\;,\;$-0.054$\;]$\tabularnewline
\hline
{\bfseries Extrapolation}&&&&&&&&\tabularnewline
~~$\tau_{\ell}(-650)$&0.191&&&&&&0.001&$[\;$0.080$\;,\;$0.336$\;]$\tabularnewline
\hline
\end{tabular}

%% file: inputs/table_results_trendsgl.txt
\begin{tabular}{lcc}
\hline\hline
\multicolumn{1}{l}{}&\multicolumn{1}{c}{Estimate}&\multicolumn{1}{c}{p-value}\tabularnewline
\hline
Constant&5.534&0.159\tabularnewline
$Score$&0.010&0.220\tabularnewline
$Score^2$&0.000&0.245\tabularnewline
$\1(C=\H)$&5.732&0.779\tabularnewline
$\1(C=\H)\times Score$&0.012&0.790\tabularnewline
$\1(C=\H)\times Score^2$&0.000&0.795\tabularnewline
$N$&257&\tabularnewline
$F$-test&&0.919\tabularnewline
\hline
\end{tabular}

%% file: inputs/table_results_trendsloc.txt
\begin{tabular}{lrcrccc}
\hline\hline
\multicolumn{1}{l}{\bfseries }&\multicolumn{1}{c}{\bfseries }&\multicolumn{1}{c}{\bfseries }&\multicolumn{1}{c}{\bfseries }&\multicolumn{1}{c}{\bfseries }&\multicolumn{2}{c}{\bfseries Robust BC Inference}\tabularnewline
\cline{6-7}
\multicolumn{1}{l}{}&\multicolumn{1}{c}{Estimate}&\multicolumn{1}{c}{}&\multicolumn{1}{c}{Bw}&\multicolumn{1}{c}{}&\multicolumn{1}{c}{p-value}&\multicolumn{1}{c}{95\%CI}\tabularnewline
\hline
$\mu^{(1)}_{\ell}(\ell)$&$$-0.00025$$&&58.9&&0.986&$[\;$-0.0179\;,\;0.0176$\;]$\tabularnewline
$\mu^{(1)}_{\H}(\ell)$&$$0.00042$$&&154.6&&0.977&$[\;$-0.0015\;,\;0.0014$\;]$\tabularnewline
Difference&$$-0.00066$$&&&&0.988&$[\;$-0.0180\;,\;0.0177$\;]$\tabularnewline
\hline
\end{tabular}

%% file: inputs/table_results_simul.txt
\begin{tabular}{lrcccccc}
\hline\hline
\multicolumn{1}{l}{\bfseries }&\multicolumn{1}{c}{\bfseries }&\multicolumn{1}{c}{\bfseries }&\multicolumn{3}{c}{\bfseries Point Estimation}&\multicolumn{1}{c}{\bfseries }&\multicolumn{1}{c}{\bfseries RBC Inference}\tabularnewline
\cline{4-6} \cline{8-8}
\multicolumn{1}{l}{}&\multicolumn{1}{c}{Eff. $N$}&\multicolumn{1}{c}{}&\multicolumn{1}{c}{Bias}&\multicolumn{1}{c}{Var}&\multicolumn{1}{c}{RMSE}&\multicolumn{1}{c}{}&\multicolumn{1}{c}{Cov.(95\%)}\tabularnewline
\hline
{\bfseries Small $N$}&&&&&&&\tabularnewline
~~$\mu_{\ell}(\ell)$&23.4&&0.001&0.0213&0.146&&0.890\tabularnewline
~~$\mu_{\ell}(\bar{x})$&78.0&&-0.007&0.0015&0.039&&0.937\tabularnewline
~~$\mu_{\H}(\ell)$&140.9&&-0.001&0.0008&0.029&&0.945\tabularnewline
~~$\mu_{\H}(\bar{x})$&101.6&&-0.010&0.0012&0.036&&0.937\tabularnewline
~~$\tau_{\ell}(\bar{x})$&&&0.002&0.0247&0.157&&0.905\tabularnewline
\hline
{\bfseries Moderate $N$}&&&&&&&\tabularnewline
~~$\mu_{\ell}(\ell)$&43.0&&-0.001&0.0134&0.116&&0.905\tabularnewline
~~$\mu_{\ell}(\bar{x})$&136.9&&-0.005&0.0008&0.029&&0.945\tabularnewline
~~$\mu_{\H}(\ell)$&279.4&&-0.000&0.0004&0.020&&0.950\tabularnewline
~~$\mu_{\H}(\bar{x})$&213.7&&-0.012&0.0005&0.026&&0.949\tabularnewline
~~$\tau_{\ell}(\bar{x})$&&&0.008&0.0150&0.123&&0.917\tabularnewline
\hline
{\bfseries Large $N$}&&&&&&&\tabularnewline
~~$\mu_{\ell}(\ell)$&108.6&&0.001&0.0050&0.071&&0.933\tabularnewline
~~$\mu_{\ell}(\bar{x})$&288.0&&-0.003&0.0004&0.020&&0.945\tabularnewline
~~$\mu_{\H}(\ell)$&681.9&&-0.000&0.0002&0.013&&0.953\tabularnewline
~~$\mu_{\H}(\bar{x})$&517.5&&-0.012&0.0002&0.019&&0.951\tabularnewline
~~$\tau_{\ell}(\bar{x})$&&&0.007&0.0058&0.076&&0.939\tabularnewline
\hline
\end{tabular}

%% file: inputs/table_results_650_short.txt
\begin{tabular}{lrcrcrcrc}
\hline\hline
\multicolumn{1}{l}{\bfseries }&\multicolumn{1}{c}{\bfseries }&\multicolumn{1}{c}{\bfseries }&\multicolumn{1}{c}{\bfseries }&\multicolumn{1}{c}{\bfseries }&\multicolumn{1}{c}{\bfseries }&\multicolumn{1}{c}{\bfseries }&\multicolumn{2}{c}{\bfseries Robust BC Inference}\tabularnewline
\cline{8-9}
\multicolumn{1}{l}{}&\multicolumn{1}{c}{Estimate}&\multicolumn{1}{c}{}&\multicolumn{1}{c}{Bw}&\multicolumn{1}{c}{}&\multicolumn{1}{c}{Eff. $N$}&\multicolumn{1}{c}{}&\multicolumn{1}{c}{p-value}&\multicolumn{1}{c}{95\% CI}\tabularnewline
\hline
{\bfseries RD effects}&&&&&&&&\tabularnewline
~~$C = -850$&0.061&&85.0&&85&&0.282&$[\;$-0.052$\;,\;$0.179$\;]$\tabularnewline
~~$C = -571$&0.169&&136.3&&133&&0.103&$[\;$-0.039$\;,\;$0.428$\;]$\tabularnewline
~~Weighted&0.121&&&&218&&0.056&$[\;$-0.003$\;,\;$0.274$\;]$\tabularnewline
~~Pooled&0.073&&161.5&&307&&0.254&$[\;$-0.046$\;,\;$0.173$\;]$\tabularnewline
\hline
{\bfseries Naive difference}&&&&&&&&\tabularnewline
~~$\mu_{\ell}(-650)$&0.728&&239.4&&440&&&\tabularnewline
~~$\mu_h(-650)$&0.706&&131.2&&202&&&\tabularnewline
~~Difference&0.021&&&&&&0.634&$[\;$-0.050$\;,\;$0.081$\;]$\tabularnewline
\hline
{\bfseries Bias}&&&&&&&&\tabularnewline
~~$\mu_{\ell}(-850)$&0.560&&58.0&&57&&&\tabularnewline
~~$\mu_h(-850)$&0.667&&144.2&&230&&&\tabularnewline
~~Difference&-0.106&&&&&&0.017&$[\;$-0.252$\;,\;$-0.025$\;]$\tabularnewline
\hline
{\bfseries Extrapolation}&&&&&&&&\tabularnewline
~~$\tau_{\ell}(-650)$&0.128&&&&&&0.022&$[\;$0.022$\;,\;$0.286$\;]$\tabularnewline
\hline
\end{tabular}

%% file: inputs/table_results_LR.txt
\begin{tabular}{lccccccccccc}
\hline\hline
\multicolumn{1}{l}{\bfseries }&\multicolumn{1}{c}{\bfseries }&\multicolumn{1}{c}{\bfseries }&\multicolumn{1}{c}{\bfseries }&\multicolumn{1}{c}{\bfseries }&\multicolumn{3}{c}{\bfseries Constant}&\multicolumn{1}{c}{\bfseries }&\multicolumn{3}{c}{\bfseries Linear}\tabularnewline
\cline{6-8} \cline{10-12}
\multicolumn{1}{l}{}&\multicolumn{1}{c}{Window}&\multicolumn{1}{c}{}&\multicolumn{1}{c}{Eff. $N$}&\multicolumn{1}{c}{}&\multicolumn{1}{c}{Estimate}&\multicolumn{1}{c}{Fisher p}&\multicolumn{1}{c}{Neyman p}&\multicolumn{1}{c}{}&\multicolumn{1}{c}{Estimate}&\multicolumn{1}{c}{Fisher p}&\multicolumn{1}{c}{Neyman p}\tabularnewline
\hline
{\bfseries $X_i\in\mathcal{N}_{\ell}$}&&&&&&&&&&&\tabularnewline
~~$\bar{Y}_{\ell}(0,\ell)$&$[$-900$\;,\;$-850$)$&&50&&0.502&&&&0.527&&\tabularnewline
~~$\bar{Y}_{\H}(0,\ell)$&$[$-881$\;,\;$-817$]$&&50&&0.706&&&&0.707&&\tabularnewline
~~$\Delta$&&&100&&-0.204&0.000&0.000&&-0.180&0.000&0.021\tabularnewline
\hline
{\bfseries $X_i\in\mathcal{N}_{\bar{x}}$}&&&&&&&&&&&\tabularnewline
~~$\bar{Y}_{\ell}(1,\bar{x})$&$[$-675$\;,\;$-626$]$&&50&&0.760&&&&0.759&&\tabularnewline
~~$\bar{Y}_{\H}(0,\bar{x})$&$[$-675$\;,\;$-625$]$&&50&&0.743&&&&0.743&&\tabularnewline
~~Diff&&&100&&0.017&0.702&0.698&&0.016&0.718&0.716\tabularnewline
\hline
{\bfseries }&&&&&&&&&&&\tabularnewline
~~$\hat{\tau}_{\mathsf{LR}}$&&&&&0.220&0.042&0.001&&0.196&0.037&0.029\tabularnewline
\hline
\end{tabular}